\documentclass[11pt,leqno]{article}
\usepackage{latexsym}
\usepackage{epsfig}
\usepackage{color}
\usepackage{amssymb,amsmath}

\usepackage[noline,nofillcomment,noend,noresetcount,boxed,figure]{algorithm2e}

\numberwithin{equation}{section}

\newtheorem{theorem}{Theorem}[section]
\newtheorem{lemma}[theorem]{Lemma}

\newtheorem{corollary}[theorem]{Corollary}
\newtheorem{example}{Example}[section]
\def\remark #1{\noindent{\bf Remark:} #1\\}

\long\def\claim #1 #2{\bigskip\noindent{\bf Claim {#1}} {\it #2}\bigskip}
\long\def\nclaim #1 #2{\noindent{\bf Claim {#1}} {\it #2}}
\def\xclaim #1 #2{\noindent{\bf Claim {#1}} {\it #2}\bigskip}
\newenvironment{proof}{\noindent{\bf Proof:}}{\hfill $\Box $\\}

\newcommand{\ecproof}{\hfill $\diamondsuit$\\}

\renewcommand{\thetheorem}{\arabic{section}.\arabic{theorem}}
\def\sqr#1#2{{\vcenter{\vbox{\hrule height .#2pt
              \hbox{\vrule width .#2pt height#1pt \kern#1pt
              \vrule width .#2pt} \hrule height .#2pt}}}}

\def\ncas #1 {\noindent {\bf Case #1.}\ }

\def\bipart #1 #2{\bigskip \noindent {\bf #1} {\it #2}}
\def\xbipart #1 #2{\noindent {\bf #1} {\it #2}}
\def\iipart #1 #2{\bigskip \noindent {\it #1} {\it #2}}
\def\xiipart #1 #2{\noindent {\it #1} {\it #2}}
\def\brpart #1 #2{\bigskip \noindent {\bf #1} {#2}}
\def\xbrpart #1 #2{\noindent {\bf #1} {#2}}
\def\irpart #1 #2{\bigskip \noindent {\it #1} {#2}}
\def\xirpart #1 #2{\noindent {\it #1} {#2}}

\def\o {\overline}
\def\nil{\epsilon}

\def\case #1{\bigskip\noindent{{\bf Case} {\em #1}:}}
\def\subcase #1{\bigskip\noindent{{\bf Subcase} {\em #1}:}}
\def\numcase #1 #2{\bigskip\noindent{{\bf Case #1} {\em #2}:}}

\def\obs #1 {\bigskip\noindent{\bf Observation #1: }} 

\begin{document}
\title{A Data Structure for Nearest Common Ancestors with Linking%
\thanks{A preliminary version of results in this paper appeared in 
{\em Proc.~1st Annual ACM-SIAM Symp.~on Disc.~Algorithms}, 
1990 \cite{G90}.} 
\author{Harold N.~Gabow%
\thanks{Department of Computer Science, University of Colorado at Boulder,
Boulder, Colorado 80309-0430, USA. 
Research supported in part by NSF Grant No. CCR-8815636.
E-mail: {\tt hal@cs.colorado.edu} 
}
}
}
\date{December 18, 2014; revised November 18, 2016}
\maketitle
\def\today{\ifcase\month\or
January\or February\or March\or April\or May\or June\or
July\or August\or September\or October\or November\or December\fi
\ \number\day, \number\year}
\def\date#1.#2.{\ifcase#1\or
January\or February\or March\or April\or May\or June\or
July\or August\or September\or October\or November\or December\fi
\ #2, \number\year}
\def\ydate#1.#2.#3.{\ifcase#1\or
January\or February\or March\or April\or May\or June\or
July\or August\or September\or October\or November\or December\fi
\ #2, 199#3}
\def\nydate#1.#2.{\ifcase#1\or
January\or February\or March\or April\or May\or June\or
July\or August\or September\or October\or November\or December\fi
\ #2}
\def\doublespace{\multiply\baselineskip by3\divide\baselineskip by2%
                 \def\doublespace{}}
\def\bigdoublespace{\multiply\baselineskip by2%
                 \def\bigdoublespace{}}
\def\imp{\ifmmode {\ \Longrightarrow \ }\else{$\ \Longrightarrow \ $}\fi}
\def\rimp{\ifmmode {\ \Longleftarrow \ }\else{$\ \Longleftarrow \ $}\fi}
\def\ximp{\ifmmode {\Longrightarrow\ }\else{$\Longrightarrow\ $}\fi}
\def\xrimp{\ifmmode {\Longleftarrow\ }\else{$\Longleftarrow\ $}\fi}
\def\iff{\ifmmode {\ \Longleftrightarrow \ }\else{$\ \Longleftrightarrow \ $}\fi}
\def\xiff{\ifmmode {\Longleftrightarrow\ }\else{$\Longleftrightarrow\ $}\fi}
\def\tru{\ {\bf true}\ }
\def\fal{\ {\bf false}\ }
\def\wrt{\ {\it wrt}\ }
\def\endskip{\medskip}
\def\qed{$\Box$}
\def\qedn{\ \vrule width4pt depth-1pt height7pt }
\def\rqed{\hfill\hbox to 24 pt{\vrule width4pt depth-1pt
height7pt\hfil}\bigskip}
\def\rqedn{\hfill\hbox to 24 pt{\vrule width4pt depth-1pt height7pt\hfil}}
\def\log{\ifmmode \,{ \rm log}\,\else{\it log }\fi}
\def\con {\subseteq}
\def\pcon{\subset}
\def\firstnumstp#1 {\bigskip \noindent{\it Step} #1.\newquad}
\def\numstp#1 {\endskip\noindent{\it Step} #1.\newquad}
\def\newquad{\hskip1ex}
\def\stp#1.{\endskip
\noindent{\it #1 Step.}\newquad}
\def\firststp#1.{\bigskip
\penalty-1000
\noindent{\it #1 Step.}\newquad}
\def\cas#1 {\smallskip\noindent{\bf Case} #1.\ } 
%
%
%
\long\def\sec#1{\bigskip
\penalty-2000%
\noindent{\twelvebf #1}\par\ignorespaces\noindent\ignorespaces}
\def\aorbsec#1{\noindent{\twelvebf #1}}
\def\nsec#1{\penalty-2000%
\noindent{\bf #1\hfill\break}
\hbox to \parindent{\hfill}\ignorespaces}
\long\def\res #1. #2{\bigskip
\penalty-1000
\noindent {\bf #1.}\newquad%
#2 \bigskip}
\long\def\nres #1. #2{\bigskip
\noindent {\bf #1.}\newquad%
#2}
\def\pf{\noindent {\bf Proof.}\newquad}
\def\cont{\ifmmode\star\else$\star$\fi}
\def\+{\tabalign} 
\def\nskp{\def\bigskip{}}
\def\i{($i$) } \def\xi{($i$)}
\def\ii{($ii$) } \def\xii{($ii$)}
\def\iii{($iii$) } \def\xiii{($iii$)}
\def\iv{($iv$) } \def\xiv{($iv$)}
\def\pa{({\it a}) } \def\xpa{({\it a})} 
\def\pb{({\it b}) } \def\xpb{({\it b})}
\def\pc{({\it c}) } \def\xpc{({\it c})}
\def\hi{\hskip20pt\i} \def\hii{\hskip20pt\ii} \def\hiii{\hskip20pt\iii}
\def\ha{\hskip20pt\pa} \def\hb{\hskip20pt\pb} \def\hc{\hskip20pt\pc}
\def\tran{{\buildrel*\over\to}}
\def\n{\rlap{$\>/$}}
\def\({{\rm(}} \def\){{\rm)}}
\def\c#1{\lceil {#1} \rceil}
\def\f#1{\lfloor {#1} \rfloor}
\long\def\boxit#1{\vtop{\hrule
\hbox{\vrule\quad\vtop{\vskip5pt\hbox{#1}\vskip5pt}\quad\vrule}
\hrule}} 
\def\iboxit#1{\vtop{\hrule
\hbox{\vrule\quad\vtop{\vskip5pt\hbox{{\it #1}}\vskip5pt}\quad\vrule}
\hrule}} 
\def\x{\iffalse}
\def\b{\bigskip}
\def\set #1#2{\{ #1:#2 \}}
\def\pset #1#2{( #1:#2 )}
\def\h{\hskip20pt}
\def\hi{\advance\parindent by 20pt}

\def\o{\overline} 
\def\u{\underline}
\def\opn{\hangindent=40pt\hangafter=1}
\def\h{{\hskip 20pt}}
\def\v{\vfill}
\def\hi{\advance \parindent by 20pt}
\def\d{\cdot}
\def\al.{{\it add\_leaf}}
\def\alm{{\it add\_leaf}$\,$}
\def\O{o\hbox{-}smallest}
\def\os.{\ifmmode{ \o{\cal S} }\else{$\o {\cal S}$}\fi}
\def\oP.{\ifmmode{ \o{\cal P} }\else{$\o {\cal P}$}\fi}
\def\ot.{\mathy{ \o{\cal T} }}
\def\oG{\o G}
\def\oB{\o B}
\def\oE.{\mathy{\overline E}}
\def\p(#1,#2){\ifmmode p(#1,#2) \else{$p(#1,#2)$}\fi}
\def\op(#1,#2){\ifmmode \o{p}(#1,#2) \else{$\o{p}(#1,#2)$}\fi}
\def\lb{\ifmmode \,{ \rm log}_\beta \else{\it log XX }\fi}
\def\wh{\widehat}
\def\wx.{\ifmmode \wh x \else$\wh x$\fi}
\def\wy.{\ifmmode \wh y \else$\wh y$\fi}
\def\wz.{\ifmmode \wh z \else$\wh z$\fi}
\def\wv.{\ifmmode \wh v \else$\wh v$\fi}
\def\Px.{\ifmmode \wh x \else$\wh x$\fi}
\def\Py.{\ifmmode \wh y \else$\wh y$\fi}
\def\Pz.{\ifmmode \wh z \else$\wh z$\fi}
\def\Pv.{\ifmmode \wh v \else$\wh v$\fi}
\def\Pr.{\ifmmode \wh r \else$\wh r$\fi}
\def\Pr.{\ifmmode \wh r \else$\wh r$\fi}
\def\A.{\mathy{{\cal A}}}
\def\B.{\mathy{{\cal B}}}
\def\E.{\ifmmode {{\cal E}}\else{{$\cal E$}}\fi}
\def\F.{\mathy{\cal F}}
\def\H.{\mathy{\cal H}}
\def\M.{\mathy{\cal M}}
\def\P.{\mathy{\cal P}}
\def\S.{\ifmmode {{\cal S}}\else{{$\cal S$}}\fi}
\def\mathy #1{\ifmmode {#1}\else{$#1$}\fi}
\def\pkt(#1,#2){\mathy{packet(#1,#2)}}

\def\rt{\mathy{\rho}}
\def\pt{\pi}
\long\def\example #1. #2{\bigskip \noindent{\bf Example #1.} 
{#2}\bigskip} 
\long\def\xexample #1 {\bigskip \noindent{\bf Example.} 
{#1}\bigskip}
\def\goin{\hspace{17pt}}

\begin{abstract}
Consider a forest that evolves via $link$ operations that make the
root of one tree the child of a node in another tree.  Intermixed with
$link$ operations are $nca$ operations, which return the nearest
common ancestor of two given nodes when such exists.  This paper shows
that a sequence of $m$ such $nca$ and $link$ operations on a forest of $n$
nodes can be processed on-line in time $O(m\alpha(m,n)+n)$.  This was
previously known only for a restricted type of $link$ operation.  The
special case  where a $link$ only extends a tree by
adding a new leaf occurs in Edmonds' algorithm for finding a maximum
weight matching on a general graph. Incorporating our algorithm into
the implementation of Edmonds' algorithm in \cite{G17} achieves time
$O(n(m + n\log n))$ for weighted matching, an arguably optimum
asymptotic bound ($n$ and $m$ are the number of vertices and edges,
respectively).
\end{abstract}

\def\switch{0}
\ifcase \switch 
\section {Introduction} 
\label{IntroSec}
Finding nearest common ancestors 
($nca$s) is a basic data structure operation. 
It is used to solve many  string problems via the suffix tree
\cite{CH, Gus, Smy}. It arises 
in trees that grow by adding leaves,
in the efficient implementation of algorithms for
weighted matching and its generalizations \cite{G90, G17}.
This application is the jumping-off point for this paper.

We summarize our algorithms for dynamic $nca$s, 
starting with the most general version of the problem.
Recall that 
the {\it nearest common ancestor} of two nodes $x$ and $y$ in a rooted
tree
is the ancestor of $x$ and $y$ that has greatest depth.
Consider a rooted forest $F$ that is subject to 
two operations. In both operations $x$ and $y$ are nodes of $F$.
\b

\long\def\twolines #1 #2{\vtop{\hbox{#1}\hbox{#2}}}

{

{

$link(x,y)$ -- make $y$ a child of $x$, where
$y$ is the root of a tree not containing $x$;

$nca(x,y)$ -- \twolines {return the nearest common ancestor of $x$ and $y$;} 
{if $x$ and $y$ are in different trees, return $\emptyset$.}

}
}

\b
\noindent
The problem of {\it nearest common ancestors with linking} is to
process (on-line) an arbitrary
sequence of $m$ {\it link} and {\it nca} operations, starting from
an initial forest of $n$ nodes ($m, n\ge1$).

A number of data structures have been proposed for this problem and
various special cases. See \cite{HT} for the early history, which
dates back to
the off-line algorithm  of Aho, Hopcroft and Ullman \cite{AHU}.
The problem of {\it nearest common ancestors for static trees} is when
$F$ is given initially (equivalently, all {\it links} precede
all {\it ncas}). Harel and Tarjan give an algorithm 
that answers each $nca$ query in time $O(1)$, after
$O(n)$ time to preprocess $F$ \cite{HT}.
Schieber and Vishkin
\cite{SV} simplify this approach, and
Berkman and Vishkin \cite{BV} give a yet simpler algorithm
based on Euler tours.
Harel and Tarjan
also give an algorithm for the case where {\it link} and {\it nca}
operations are intermixed but
both arguments of every {\it link} are roots
\cite{HT}. The
running time is $O(m\alpha(m,n)+n)$. 
In this paper we remove the restriction on links: 
We show that the general nearest common ancestors with linking problem
can be solved in time $O(m\alpha(m,n)+n)$ and space $O(m+n)$.
The previous best solution is using dynamic trees \cite{SlT}.
This data structure performs
each operation in time $O(\log n)$, achieving total time $O(m\log n +n)$.
This is not as fast as our algorithm, although
dynamic trees have the advantage that they can also process {\it cut}
operations. Our result is in some sense optimal: Fredman has
pointed out that the results of \cite{FS} can be extended to show
that nearest common ancestors with linking requires time
 $\Omega(m\alpha(m,n)+n)$ in the cell probe model of computation.

Dynamic $nca$s arise
in Edmonds' algorithm for weighted matching on graphs \cite{Ed, G17, S}.
The algorithm is based on the notion of 
``blossom'' --  a type of subgraph that can be contracted in 
the search for an augmenting path.
Blossoms get contracted using a trivial $nca$ computation in a search graph.
But only the ``cheapest blossom'' can be contracted in Edmonds' algorithm. 
An efficient
matching algorithm must track all possible blossoms and contract only the best.
This necessitates answering an $nca$ query for every possible blossom.
Furthermore the query is made in the search graph -- a forest 
that grows by adding new leaves as the search progresses. 

To be precise consider the following operations on a rooted tree
$T$.
$x$ is a node already in $T$, and $y$
is a new node not yet in $T$:

\b

{


\alm$(x,y)$ -- add a new leaf $y$, with parent $x$, to  $T$;

$add\_root(y)$ -- make the current root of $T$ a child of new root $y$.

}

\b

\noindent
The problem of {\it incremental-tree nearest common ancestors} is to
process (on-line) an arbitrary sequence of $n$ 
\al. and $add\_root$ operations intermixed with $m$ {\it nca}
operations, starting with an empty tree (more precisely 
$T$ starts with one node, an artificial root).
Edmonds' algorithm actually only uses \al. and {\it nca} operations.
We give an algorithm
that solves this problem in $O(m+n\log n)$ time. 
This achieves the desired time bound for Edmonds' algorithm -- 
it finds a maximum weight matching in  time 
$O(n(m+\log n))$ \cite{G90,G17}.%
\footnote{This time bound is optimal in an appropriate model of
computation \cite{G90,G17}.} 

We refine our $nca$ algorithm to use
time
$O(m+n)$. This incremental-tree algorithm
becomes the starting point of our algorithm for general {\it links}.%

After the conference version of this paper Cole and Hariharan
presented
a more powerful algorithm for the
incremental-tree $nca$ problem.
Our  algorithm answers each $nca$ query in 
worst-case time $O(1)$,
but $add$ operations are $O(1)$ only in an amortized sense.
Cole and Hariharan achieve  worst-case time $O(1)$ for all operations, even allowing
insertion of internal nodes and deletion of
nodes with $\le 1$ child \cite{CH}. The starting point of their algorithm
is a version of our approach presented in Section \ref{3.1Sec}.%
\footnote{The {\em reorganize} operation defined in Section \ref{3.1Sec} 
makes our time-bound for $add$s amortized. Cole and Hariharan
do reorganizations piecemeal in future operations, so each operation does constant work \cite{CH}.}


The model of computation throughout this paper is a random access machine 
with a
word size of $\log n$ bits. \cite{HT} gives a lower bound indicating it is unlikely that our results for nearest common ancestors
can be achieved on a pointer machine. However
it still might be possible to achieve our
results for Edmonds' algorithm on a pointer machine.

The rest of the paper is organized as follows.  
This section concludes with some terminology.
Section \ref{3.1Sec} introduces the basic idea, a generalization of 
preorder numbering of trees. It
solves the  incremental-tree $nca$ problem in
$O(m+n\log^2n)$ time.
Section \ref{3.2Sec}
improves the time bound $O(m+n\log n)$.
This is all that is needed to complete the implementation
of Edmonds' algorithm in time $O(m+n\log n)$.
(Readers interested only in the application to matching need go no further.)
The approach of Section  \ref{3.2Sec} is  extended
in the next two sections:
Section \ref{3.3Sec}
achieves linear time for the
incremental-tree nca problem.
Section \ref{3.4Sec}
achieves time $O(m\alpha(m,n)+n)$
to process $m$ $nca$ and {\it link} operations
on a set of $n$ nodes.
Some further details are presented in two appendices:
Appendix A shows how to compute the logarithms needed
in Section
\ref{3.1Sec}. Appendix B
proves simple properties
of Ackermann's function used in Section \ref{3.4Sec}.

\paragraph*{Terminology}
We use interval notation for sets of integers: for $i,j\in \mathbb{Z}$, 
$[i..j]=\set {k^{\in \mathbb{Z}}} {i\le k\le j}$. 
$\log n$ denotes logarithm to the base two.

As usual we assume a RAM machine does truncating integer division.
We compare two rational numbers $a/b$ and $c/d$
for positive integers $a,b,c,d$ by comparing $ad$ and $bc$. 
Assume that for a given integer $r\in [1..n]$ the value
$\f{\log r}$ can be computed in $O(1)$ time.
This can be done if we precompute these $n$
values and store them in a table. 
The precomputation time is $O(n)$.
More generally for a fixed rational number $\beta>1$
$\f{\log_\beta r}$ can be computed in $O(1)$ time (see Appendix \ref{LogAppendix}).

We use the following tree terminology.
Let $T$ be a tree.
$V(T)$ denotes its vertex set.
A {\it subtree} of $T$ is a connected subgraph.
The root of $T$ is denoted $\rt(T)$. Let $v$ be a node of $T$.
The {\it ancestors} of $v$ are the
nodes on the path from $v$ to $\rt(T)$. The ancestors are ordered as in this
path. This indicates how to interpret expressions like
``the first ancestor of $v$ such that''.
The {\em descendants} of $v$
are all nodes that have $v$ as an ancestor.
$T_v$ denotes the subtree of all descendants of  $v$.
The {\it parent} of $v$ is denoted $\pt(v)$.
For any function $f$ defined on nodes of a  tree, we write
$f_T$ when the tree $T$ is not clear from context
 (e.g., $\pt_T(v)$).

\def\wa{\mathy{\wh{\,a\,}}}
\def\wl{\mathy{\wh{\,l\,}}}
\def\wca{\mathy{\wh{\,c\,}}}

\def\fx.{\mathy{\overrightarrow{x}}}
\def\gx.{\mathy{\overleftarrow{x}}}

\def\fy.{\mathy{\overrightarrow{y}}}
\def\gy.{\mathy{\overleftarrow{y}}}

\def\fa.{\mathy{\overrightarrow{a}}}
\def\fc.{\mathy{\overrightarrow{c}}}

\def\fz #1.{\mathy{\overrightarrow{#1}}}
\def\gz #1.{\mathy{\overleftarrow{#1}}}

\def\wbeta.{\mathy{\wh{\,\beta\,}}}

\section{Fat preorder for dynamic trees}
\label{3.1Sec}
This section introduces the basic idea for
dynamic trees, the fat preorder numbering that generalizes
preorder numbering of trees.
It starts with an algorithm to find nearest common ancestors
on a tree that is given in advance.
Then it extends that algorithm
to solve the  incremental-tree nca problem in
$O(m+n\log^2n)$ time.

Our main auxiliary tree is the compressed tree,
so we begin by reviewing the basic definitions \cite{HT, T79}.
Let $T$ be a tree with root $\rt(T)$. The {\it size} $s(v)$ 
of a node $v$ is the number of its descendants.
(As usual a node is a descendant of itself.)
A child $w$ of $v$ is  {\it light} if $s(w)\le s(v)/2$; 
otherwise it is {\it heavy}.
Deleting each edge from a light child to its parent 
partitions the nodes of $T$ into 
paths of nonnegative length, called the {\it heavy paths of $T$}.
The highest node on a heavy path is the {\em apex} of the path;
each apex has at most one child.
(So an isolated node is considered a heavy path, and a heavy path
has only one node at each depth.)
A node is an {\it apex} if it is  not a heavy child
(e.g., $\rt(T)$).
Equivalently an apex is the highest node on its heavy path.

To generalize this consider
an arbitrary partition of $V(T)$ into a family \P.
of disjoint paths of nonnegative length.
Length 0 paths are allowed, and 
we require that each path has only one node at each depth. 
Call the highest node on each path its apex.
The {\it compressed tree for $T$ and $\cal P$}, 
$C(T,{\cal P} )$, 
has nodes $V(T)$; its  root is $\rt(T)$ and
 the parent of a node $v\ne \rt(T)$
is the first proper ancestor of $v$ that is an apex.
Any apex $v$ has 
the same descendants in 
$C(T,{\cal P} )$ and $T$, so in particular 
$s_{C(T,{\cal P} )}(v)=s_T(v)$. 
When $\cal P$ consists of the heavy paths of $T$ we call 
$C(T,{\cal P} )$ 
the {\it compressed tree \(for $T$\)}, denoted $C(T)$.
As extreme examples $C(T)$ is
$T$ if $T$ is a complete binary tree, and $C(T)$ has height 1
if $T$ is a path.

Let $T$ be an arbitrary tree and let $C$ be its compressed tree
 $C(T)$. $C$ has height $\le \f{\log n}$.
This follows from the fact
that in $C$,  the parent $v$ of  node $w$ has
$s_C(v)\ge 2s_C(w).$
(This is clear if $w$ is a light child in $T$.
If $w$ is a heavy child then $s_C(w)=1$ and $s_C(v)\ge 2$.)

For any nodes $x,y$ in the given tree $T$,
we compute $nca_T(x,y)$ by starting with the corresponding
node $nca_C(x,y)$ in the compressed tree $C$.
\cite{HT} computes
$nca_C(x,y)$
 in $O(1)$ time
by embedding $C$ in a complete binary tree $B$; 
$ncas$ in $B$ are calculated using
the binary expansion of the inorder numbers  of the nodes.
\cite{SV} uses a similar approach.  
We now present  a different strategy. It seems to
give simpler algorithms (see Section \ref{3.2Sec}).

\begin{figure}[t]
\centering
\input{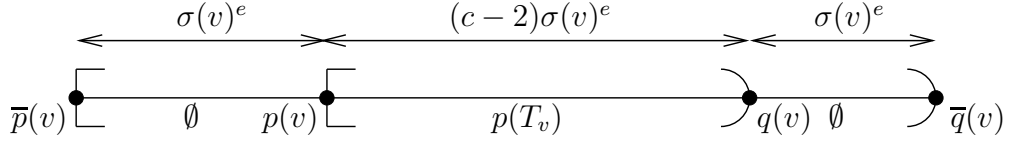}
 \caption{Fat preorder inverval for vertex $v$. The two empty intervals
guard the preorder numbers of the vertices of $T_v$, 
which are in $[p(v),q(v))$.}
 \label{FPreorderFig}
 \end{figure}

Our main tool, the fat preordering, is defined
for a tree $C$ that again generalizes the  compressed tree. 
Choose a real-valued constant $\beta>1$ and integers $e>1$, $c>2$ such that
\begin{equation}
\label{2Eqn}
{2\over \beta^{e-1}-1}\le c-2\le \beta^e.
\end{equation}
Note the left inequality is satisfied when
$\beta^{e-1}\ge 2$ and $c\ge 4$, so a convenient choice is
$e=\beta=2$ and  $c=4$.
Let $\sigma:V(C)\to [1..n]$
\footnote{Here $n$ is not necessarily
equal to $|V(C)|$. Eventually we will have
$n$ either equal to $|V(C)|$ with $\sigma=s_C, \beta=2$
or $n\le |V(C)|$ and $\sigma$ a variant of $s_C$.}
be such that
every node $v$ with child $w$ satisfies
\begin{equation}
\label{1Eqn}
\sigma(v) \ge  \beta \sigma(w).
\end{equation}
For functions $p,q,\o p,\o q: V(C) \to \mathbb{Z_+}$,
$p$ is a  {\it fat preorder numbering of $C$}
if for any node $v$, as illustrated in Fig.\ref{FPreorderFig},

\b

\i the descendants  of $v$ in $C$ are the nodes $w$ with
$p(w)\in [p(v),q(v) )$;

\ii no node $w$ has  $p(w)\in [\o p(v),p(v))\cup [  q(v),\o q(v) )$;


\iii $\o q(v)-\o p(v)= c\sigma(v)^e$ and $p(v)- \o p(v),\ 
\o q(v)-q(v)= \sigma(v)^{e}$.

\b

\noindent Note that \i is equivalent to $p$ being a
preorder numbering. Also the definition allows 
``guarding intervals'' to overlap, e.g., we may have
$\o q(w)\in [p(v),q(v))$ for $w$ not descending from $v$.
However our algorithms will maintain the intervals
$[\o p(v),\o q(v))$ as a laminar family.
%
Without loss of generality we can assume  $\o p (\rt(C))=0$,
so all $p$-numbers are in $[0,cn^e)$.
The fat preorders that we construct 
take $ \sigma$   to be  the size function
$s_C$ (for static trees) or a close variant of $s_C$ (for dynamic trees).


Given a fat preordering, the following high-level algorithm
returns $nca_C(x,y)$:

\b

{\narrower

{\parindent=0pt

Let $a$ be the first ancestor of $x$ that has
$ (c-2) \sigma(a)^{e}>|p(x)-p(y)|$.
If $a$ is an ancestor of $y$ (i.e., $p(a)\le p(y)< q(a)$)
then return $a$ else return 
$\pt_C(a)$ (the parent of $a$). 

}
}

\begin{lemma}
\label{3.1Lemma}
The  {\em nca}$_{\rm C}$ algorithm  is correct.
\end{lemma}

\begin{proof}
We first show that  any common ancestor $b$ of $x$ and $y$
satisfies 
\[ (c-2) \sigma(b)^{e}>|p(x)-p(y)|.\]
By \i the interval $ [p(b),q(b) )$
contains both $p(x)$ and $p(y)$ so its length is at least
$1+|p(x)-p(y)|$. By \ii its length
is $ \le (c-2) \sigma(b)^{e}$.
The above inequality follows.


Now to prove the lemma we need only show
that $a'=\pt_C(a)$ is a common ancestor. 
(Clearly we can assume $a$ is not the root.)
This amounts to showing
$a'$ is an ancestor of
$y$. By \i -- \iii
a nondescendant of $a'$ and a descendant of $a'$ differ in $p$-number by more
than $\sigma(a')^e$. Using (\ref{1Eqn}) and the right inequality of 
(\ref{2Eqn}),
$\sigma(a')^e\ge \beta^e \sigma(a)^e\ge (c-2) \sigma(a)^{e}>|p(x)-p(y)|$.
Since $x$ descends from $a'$, so must  $y$.
\end{proof}

We implement the high-level algorithm 
using the following data structure. Each vertex $x$
stores an {\it ancestor table}, 
$ancestor_x[0..\f{\lb cn^e}]$, where
\[ancestor_x[i] \text{ is the last ancestor $b$ of $x$ 
that has $(c-2) \sigma(b)^{e}< \beta^i$.}\]
If no such ancestor exists, i.e.,
$(c-2) \sigma(x)^{e}\ge \beta^i$, then
 the
entry is $\nil$.

Before implementing the high-level algorithm we note that computing
$nca$s using the compressed tree (and
other auxiliary trees that we shall see)
requires more than just the $nca$ node.
Fix any tree and consider nodes $x,y$. Let $a=nca(x,y)$.
For $z=x,y$, let
$a_z$ be the ancestor of  $z$ immediately preceding $a$;
if $a=z$ then take $a_z=a$.
Define $ca(x,y)$, the {\it characteristic ancestors of $x$ and $y$}, as
the ordered triplet $(a,a_x,a_y)$. Our
$nca$ algorithms actually compute $ca$.

We now implement the high-level $nca$ algorithm in $C$, in fact finding
all the characteristic ancestors, in $O(1)$ time.
We seek $nca_C(x,y)$
and the characteristic ancestor $a_x$;
$a_y$ is symmetric.

\bigskip

It suffices to find the ancestor $a$
of the $nca_C$ algorithm, plus the ancestor of $x$ that precedes
$a$ 
if $a\ne x$. (If $a=x$ then either 
$a= nca_C(x,y)=x$ giving  $a_x=x$, or 
$nca_C(x,y)=\pt_C(a)$ again giving $a_x=a=x$.)

Let $i=  \f{\lb |p(x)-p(y)| }$. 
Clearly the first ancestor $w$ of $x$ that has
\[(c-2) \sigma(w)^{e}\ge \beta^i\]
is either $a$ or a descendant of $a$ (possibly $w=a$).
Find $w$ 
as follows. Let $v= ancestor_x[i]$.
$v\ne\rt(C)$ since $(c-2)\sigma(\rt(C))^e > |p(x)-p(y)|\ge \beta^i$,
where the first inequality holds for arbitrary $x,y\in V$.
If $v\ne \nil$ then $w=\pt(v)$.
If $v= \nil$ then $w=x$.

First suppose  $w\ne\rt(C)$.
For $w'=\pt(w)$,
(\ref{1Eqn}) and $e\ge 1$ show $(c-2) \sigma(w')^{e}\ge \beta^{i+1}$.
Since $\beta^{i+1}> |p(x)-p(y)| $, 
the desired $a$ is either $w$ or $w'$.
This also holds if $w=\rt(C)$, since that makes $a=w$.
If $a=w'$ the ancestor preceding $a$ is $w$.
If $a=w$ and  $v\ne \nil$ the desired ancestor is $v$.
In the remaining case  $a=w=x$, so
the ancestor preceding $a$ is not needed.

\bigskip

The time for this procedure is $O(1)$.
(The value $i$ is computed as in Appendix \ref{LogAppendix}.) 
We apply the procedure twice (for $a_x$ and $a_y$).
Thus we have shown how to compute
$ca_C(x,y)$ in
$O(1)$ time.

%

Now let $C$ be a tree on $n$ nodes satisfying
(\ref{1Eqn}) for $\sigma=s=s_C$.
(An example is the compressed tree, with $\beta=2$.)
We show that a fat preordering of  $C$ exists
and can be constructed in $O(n)$ time.
We use a recursive
 numbering procedure. It traverses $C$ top-down.
When visiting a node $u$, $u$ will have already been assigned an
interval $[\o p(u), \o q(u))$ with $\o q(u)-\o p(u)=cs(u)^e$.
Initially assign $\rt(C)$ the interval $[0,cn^e)$.
Each child of $u$ will get the leftmost possible
interval, i.e., the intervals of $u$'s children will form a
partition of an interval beginning at $p(u)+1$.
To visit $u$ execute the following procedure:

\b

{\narrower

{\parindent=0pt

Assign $p(u)\gets \o p(u)+ s(u) ^e$ and
$q(u)\gets \o q(u)-s(u)^e$. Then assign intervals to  the children of $u$,
starting at $p(v)+1$, as follows: 

\smallskip

For each child $v$ of $u$,
assign the next interval 
in $[p(u),q(u))$
of length $cs(v)^e$ 
to $v$, and then visit $v$.

}}

\begin{lemma}
\label{3.2Lemma}
The numbering algorithm 
gives a valid fat preordering with $\sigma=s_C$ when
$C$ is a tree 
satisfying (\ref{1Eqn}) for $\sigma=s_C$.
The time is $O(n)$ for $n=|V(C)|$.
\end{lemma}

\begin{proof}
It is clear that the algorithm achieves
properties \i -- \iii of fat preorder,
and it runs in $O(n)$ time. We must
show that the intervals assigned by $u$ 
all fit into the interval $[\o p(u),\o q(u))$  given to $u$. 
For $u$ a leaf this holds since 
the interval's size is
$c\ge  3$. 
Assume $u$ is an interior node and let $U$ denote the set of children of $u$. 
Starting with the relation
$s(u)=1+\sum_{v \in U} s(v)$,
multiply by $s(u)^{e-1}$ and use (\ref{1Eqn}) 
(and its consequence $s(u)\ge\beta$) to obtain
$s(u)^{e}\ge \beta^{e-1}+\beta^{e-1}\sum
_{v\in U} s(v)^e$. 
This implies
 $cs(u)^{e}/\beta^{e-1} \ge 1+\sum_{v\in U} cs(v)^e$.
The right-hand side is
the total size of intervals assigned in $[p(u),q(u) )$
(the term 1 accounts for the number $p(u)$).
Since $[p(u),q(u) )$ has length $(c-2)s(u)^e$ it suffices to have
$c-2\ge c/\beta^{e-1}$. This is equivalent to the left inequality of (\ref{2Eqn}).
\end{proof}

After the numbering the tree in fat preorder
we construct the ancestor tables:
For every node $x$ we  find successive entries of
$ancestor_x[i]$  by traversing the path from $x$ to $\rho(C)$.
The time and space for this
is $O(n \log n)$ and dominates the entire algorithm.

The last important step in the $nca$ algorithm is a procedure
(due to \cite{HT}) that computes characteristic ancestors in $T$ from
those in the compressed tree $C(T)$. 
To state it let $C=C(T,\cal P)$ for an {\em arbitrary} 
set of paths $\cal P$. 
Suppose the characteristic ancestors $ca_C(x,y)=(c,c_x,c_y)$ are known
and we seek the characteristic ancestors $ca_T(x,y)=(a,a_x,a_y)$.

\bigskip

Let $P$ be the path of $\cal P$ with apex $c$. 
The definition of $C$ implies that
$a=nca_T(x,y)$ is the first common ancestor of $x$ and $y$ on $P$.
For $z\in \{x,y\}$ let $b_z$ denote the first ancestor of $c_z$
on $P$, i.e., $b_z$ is $c_z$ or $\pt_T(c_z)$.
Then  $a$  is  
the shallower vertex of $b_x$ and $b_y$.

Next we show how to find $a_x$ (the same procedure applies to $a_y$).
Consider three cases.

\case{$a\ne b_x$}
This makes the predecessor $a^-$ of $a$ on $P$  an ancestor of 
$b_x$ (possibly $a^-=b_x$). Thus $a^-$ is an ancestor of
$x$. Clearly this makes $a_x=a^-$.

\case{$a=b_x \ne c_x$}
This makes
$b_x=\pt_T(c_x)$. Combining equations gives  $a=b_x=\pt_T(c_x)$.
So by definition $a_x=c_x$.

\case{$a=b_x =c_x$}
We will show
\[c_x=x.\]
Combining this with the case definition gives $a=x$. 
So the definition of $ca_T$ shows $a_x=x$.
(This also makes $a_x=c_x$ as in the previous case.)

If $c_x$ is a leaf of $C$
the displayed equation follows since $c_x$ is an ancestor of $x$.
If $c_x$ is not a leaf of $C$
then $c_x$ is the unique vertex of $P$ that is nonleaf,
i.e., $c_x=c$. The definition of $ca_C$ shows
$x=c$. Combining gives $c_x=c=x$. 

\b

\bigskip

Putting these pieces together gives 
our algorithm for $nca$ queries on a static tree.
Let us summarize the algorithm. A preprocessing step computes the 
compressed tree $C=C(T)$. It is numbered in fat preorder ($\beta=2$).
In addition the order of nodes in each heavy path is recorded
(so we can find $a^-$ in the first case above).
The ancestor tables for $C$ are constructed.
The query algorithm computes characteristic ancestors, by finding
$ca_C(x,y)$ and using it to find $ca_T(x,y)$.

\begin{lemma}
\label {3.3Lemma} 
A
tree $T$ with $n$ nodes can be preprocessed using $O(n\log n)$ time and space
so that  {\em ca} queries can be answered in $O(1)$ time.\hfill$\Box$
\end{lemma}

Note that the preprocessing time and the space are both $O(n)$ except for the
resources needed to compute and store the ancestor tables.

\subsection*{Incremental trees}
We extend these ideas to trees that grow by \al. operations.
It is then to easy to complete the incremental tree data structure by
incorporating
$add\_root$ operations.

We start by presenting the high-level strategy.
Then we give the data structure and  detailed algorithm,
and finally prove that it works correctly.

We use a dynamic version $D$ of the compressed tree,
maintaining a fat preordering and computing $ca$s as before.
In more detail,
$D$ is maintained to be $C(T,\P.)$ for a time-varying 
collection
of  paths $\cal P$ that always partitions  $V(T)$.
\al. makes the new leaf a singleton path of \P.. 
The algorithm to maintain $D$ is based on
this operation: Let $v$ be an apex of $D$ (i.e., a shallowest
vertex on some path of \P.). Thus $V(D_v)=V(T_v)$.
To {\it recompress $v$}
means to replace $D_v$ in $D$ by $C(T_v)$.
As usual $C(T_v)$ is defined using the heavy paths of $T_v$, and 
the recompression changes \P.
accordingly.
Each node of $T_v$ gets {\it reorganized} in this recompression.

Recompressing $v$ updates the fat preordering of $D$ as follows.
Let $s$ be the size function on the recompressed subtree $D_v=C(T_v)$. 
The fat preordering will take  $\sigma$ to be $s$.
The other parameters of the ordering are specified below.
If $v=\rt(D)$ the recompression
uses the (static) fat preordering algorithm to assign new numbers 
to the nodes of $D_v$ in the interval $[0,cs(v)^e)$.
If $v\ne\rt(D)$ let $u$ be the parent $u=\pt_D(v)$.
Let $\o Q(u)$ be 
the currently largest value $\o q(z)$  for a child $z$ of $u$ in $D$.
The {\it expansion interval} for  $u$  is  $[\o Q(u),q(u) )$.
Clearly no numbers have been assigned in this interval.
Use the fat preordering algorithm to assign new numbers to the nodes
of $T_v$, in the interval $[\o Q(u), \o Q(u)+cs(v)^e )$.
This 
updates $\o Q(u)$ to $\o q(v)$,
decreasing the size of $u$'s expansion interval
by $cs(v)^e $.
The old interval 
for $v$, $[\o p(v),\o q(v) )$, will no longer be used, in effect it is
discarded.

The last part of the high-level description is based on a
parameter $\alpha$, $3/2>\alpha>1$.
For any node $u$ let $s(u)$ denote its current size in $D$ and let
$\sigma(u)$ 
denote the $\sigma$-value for $u$ in the current fat preordering (i.e.,
$u$'s interval $[\o p (u),\o q(u))$ has size $c\sigma(u)^e$).
$s(u)$ equals $\sigma(u)$ plus the number of descendants that $u$ has gained
since its reorganization.
$D$ is  maintained to always have 
\[s(u)<\alpha \sigma(u)\] for every node
$u$.

We turn to the data structure. In $D$
the data structure maintains the values 
of $\o p, p, q , \o q, 
\o Q, s$ and $\sigma$ for every vertex $u$. 
It also marks the apexes. The $D$ tree is represented by parent pointers.
The
tree $T$ is represented by children lists (i.e., each node has a list of its
children).
Also each path of \P. is recorded (for the $ca$ algorithm).

Now we give the detailed algorithms.
The $ca$ algorithm is the same as  the static case.
\alm$(x,y)$
proceeds as follows:

\b

{\narrower

{\parindent=0pt

Add $y$ to the list of children of $x$ in $T$. 
Make $y$ a singleton path of $\cal P$ by marking it an apex and setting
$\pt_D(y)$ to $x$ if $x$ is an apex, else $\pt_D(x)$.
Increase $s(a)$ by 1 for each ancestor $a$ of $y$ in $D$.
Let $v$ be the last ancestor of $y$ that 
now has $s(v)\ge\alpha \sigma(v)$. 
(This condition holds for $v=y$ by convention.)
Recompress $v$.
(Start by computing the values $s_T(z)$ for $z\in T_v$.)
Then construct the new
ancestor table for each  node of $T_v$.

}

}

\b

%
%

Note that if $v=y$ in this algorithm, recompressing $v$ 
does not change $D$ but assigns $v$ its
fat preorder values.

Finally we give the parameters for the fat preorder.
As we shall show, they are selected so the above strategy can be carried out,
in particular  expansion intervals are large enough.
Starting with the above parameter $\alpha\in (1,3/2)$, we will use
\[\beta= {2\over 2\alpha-1}\]  
as the constant of (\ref{1Eqn})
and in addition to inequality (\ref{2Eqn}) we require
%
\begin{equation}
\label{3Eqn}
c {(\alpha-1/2)^e +1/2^e\over1-1/\alpha^e}
\le  c-2. 
\end{equation}
Notice $\alpha\in (1,3/2)$ implies the fraction of the left-hand side
approaches 0 as $e\to \infty$ so this inequality 
can always be achieved.
For example take $\alpha=6/5$, $\beta=10/7$, $e=4$, $c=5$.
(Then (\ref{2Eqn}) amounts to
$1.1\le 3\le 4.1$ and (\ref{3Eqn}) amounts to $2.93\le 3$.)
The fat preorder must satisfy the defining properties \i -- \iii
for these parameters.

To show the algorithm is correct we must first ensure that expansion intervals
are large enough. More precisely 
consider an apex $u$ of $D$ that has just been reorganized.
The algorithm adds $<(\alpha -1)\sigma(u)$ descendants of $u$ in $D$
before recompressing $D_u$. The additions may cause  
various children $v$ of $u$
to get recompressed and thus assigned new intervals in $[p(u),q(u))$.
We must show the total length of all intervals 
ever assigned to children 
of 
$u$ is $<q(u)-p(u)=(c-2)\sigma(u)^e$. Here strict inequality accounts for the fact
that the integer $p(u)$ is assigned to $u$.
Also note that the ``total length'' includes both the original intervals
assigned when $u$ is reorganized, and the new intervals.

We will use the following notation.
We continue to let $\sigma(u)$ denote its value when $u$ is reorganized.
Let $v$ be a child of $u$ in $D$. 
If $v$ is an apex it may get recompressed some number of times, say $\ell$ times.
Let $\sigma_i(v)$, $i=0,\ldots,\ell$ be the various values of $\sigma(v)$.
For example
$\sigma_0(v)$ is the value of $\sigma(v)$ when $u$ is reorganized.
For $i\ge 1$, 
\begin{equation}
\label{ReorganizeEqn}
\sigma_i(v)\ge \alpha \sigma_{i-1}(v).
\end{equation}
(If $v$ is not an apex then $\ell=0$ and $\sigma_0(v)=1$.)

\begin{lemma}
From when $u$ is reorganized  until its next reorganiation,
the total length of all intervals 
ever assigned to children of $u$ is $<q(u)-p(u)$.
\end{lemma}

\begin{proof}
For any child $v$ of $u$,
\eqref{ReorganizeEqn} implies $\sigma_\ell(v)\ge \alpha^i\sigma_{\ell-i}(v)$
for every $i=0,\ldots, \ell$. 
Thus the total size of all intervals ever
assigned to $v$ is strictly less than
\[c\sigma_\ell(v)^e (1+ (1/\alpha)^e+(1/\alpha)^{2e}+\ldots\ )
= { c\sigma_\ell(v)^e  \over  1-1/\alpha^e }.\]
Obviously this holds for children with $\ell=0$ too.
So the total size of intervals ever assigned to children of $u$
is strictly less than
\[S={c\sum_{v\in U} \sigma_\ell(v)^e  \over  1-1/\alpha^e }.\]

We can assume $u$ has at least two children when it is initially reorganized.
(If $u$ starts with $\le 1$ child, $u$ gets reorganized as soon
as it gains a child, since 
$\alpha\sigma(u) < (3/2)\sigma(u) \le \sigma(u)+1$.) 
Right after $u$ was reorganized, every child $v$ 
had $\sigma_0(v)=s(v)\le s(u)/2=\sigma(u)/2$ descendants in $D$. 
$u$ gets $<(\alpha-1)\sigma(u)$ new descendants before reorganization.
The sum of all $\sigma_\ell(v)$ values is less than $\alpha \sigma(u)$.
Since $e>1$ simple calculus shows $S$ is maximized when 
$u$ starts with exactly two children, each with 
$\le \sigma(u)/2$ children, and every new node descends from
the same initial child.
(For any initial configuration, $S$ is maximized when all new 
nodes descend from the child that starts with the greatest value $\sigma_0(v)$.
This value in turn is maximized when there are only two descendants, 
and each starts with  $\le \sigma(u)/2$ descendants.)
Thus the maximum value of $S$ is
\[c\sigma(u)^e { (\alpha-1/2)^e + 1/2^e   \over
1-1/\alpha^e  }.\]
 The left inequality
of (\ref{3Eqn}) 
implies
$S\le (c-2)\sigma(u)^e$ as desired.
\end{proof}

Now we complete the correctness proof.

\begin{lemma}
The {\em add\_leaf} and {\em ca}  algorithms are correct.
\end{lemma}

\begin{proof}
We start by verifying (\ref{1Eqn}),
i.e., at any time when $v$ is a child of $u$ in $D$,
the current $\sigma$ function satisfies
$\sigma(u)\ge \beta \sigma(v)$. 
Immediately after $u$ was last reorganized we have 
$\sigma(u)=s(u)\ge 2s(v)=2\sigma(v)$. This inequality holds
even if $v$ has not been added by \alm, since we take
$\sigma(v)=0$. 
After the reorganization $\sigma(u)$ does not change,
and 
$u$ gets less than $(\alpha-1)\sigma(u)$ new descendants. 
Thus at any time $s(v)\le \sigma(u)/2+
 (\alpha-1)\sigma(u)= (\alpha-1/2)\sigma(u)$.
We always have $\sigma(v)\le s(v)$. Thus
$\sigma(v)/ (\alpha-1/2)\le \sigma(u)$,
i.e., $\beta\sigma(v)\le \sigma(u)$.


The defining
properties \xi--\iii 
of fat preorder numbers hold since recompression uses the
static preorder numbering algorithm.
(Since $n$ is nondecreasing, the current $\sigma$ always has values
$\le n$.)
The rest of the data structure
consists of the $ancestor$ tables
and the orderings of the paths of  \P..
If $x\in T_v$, $ancestor_x$ is
constructed  in entirety in the recompression. 
If $x\notin T_v$ the recompression does not change
 $ancestor_x$.
This table
remains valid since no ancestor $b$ of
$x$ is in $T_v$, so $\sigma(b)$ does not change.
Similarly a path of \P. with an apex not in
$T_v$ is vertex-disjoint from $T_v$. So it does
 not change when $v$ is recompressed and
 the data structure representing it remains valid.

The $ca$ algorithm 
works correctly since the data structure is correct.
\end{proof}

\begin{lemma}
\label{3.5Lemma} 
The nearest common ancestors problem with \alm\ and {\em ca}
operations can be solved in $O(m+n\log^2n)$ time and $O(n\log n)$
space.
\end{lemma}

\begin{proof}
A $ca$ operation uses $O(1)$ time, so the $ca$s use
$O(m)$ time total. 

In \alm$(x,y)$ examining each ancestor of $y$ uses $O(\log n)$ time.
Recompressing a node $v$ uses $O(s(v))$ time for all processing
except constructing the new ancestor tables, which 
uses $O(s(v)\log n)$ time.
Hence the time for an \al. operation is $O(s(v)\log n)$.
The algorithm's recompression strategy implies
$s(v)<1+\alpha \sigma(v)\le (1+\alpha) \sigma(v)$.
So the time is \[O(\sigma(v)\log n).\]

When $v$ is recompressed
%
$\ge(\alpha-1)\sigma(v)$
descendants $y'$ of $v$ have been added since the last reorganization of $v$.
Charge the above time $O(\sigma(v)\log n)$
to these new descendants 
$y'$, at the rate of
$O(\log n)$ per node.
Since $\alpha>1$ this accounts for the time recompressing $v$.

A given node $y'$ gets charged at most once from a node $v$
that was an ancestor of $y'$ when \al. added $y'$. 
(In proof,
recompressing $v$ reorganizes every new ancestor $w$ of $y'$.
So $y'$ will not be charged from $w$. In other words after $y'$ is charged from
$v$ it is  only charged from proper ancestors of $v$.)
Thus (\ref{1Eqn}) implies any $y'$ is charged $\le \lb cn^e$ times total. 
So
the total time 
charged 
to $y'$ is 
$O(\log^2 n)$.
The time bound follows.

For the space bound
note that the ancestor tables use space $O(n\log n)$.
The remaining space (as specified in the data structure for $D$) 
is linear.
\end{proof}
 
The lemma does not require 
$n$ (the number of \alm\ operations) to be known in advance. 
(This is the case in most of our applications of this algorithm, 
although not the implementation of Edmonds' algorithm.)
The timing analysis still applies verbatim.
The use of ancestor tables necessitates a storage management system
to achieve
the lemma's space bound. We use a standard doubling strategy,
which we make precise in the following lemma.

Consider a collection of arrays $A_i[1..n_i]$ that grow by adding entries.
Each such operation enlarges $A_i$ with a new entry, i.e., 
$n_i$ increases by 1 and the contents of old entries in $A_i$ do not change.
We implement this operation 
by allocating space for all arrays $A_i$ sequentially within a 
larger array $S$.
When the current version of $A_i$ in $S$  becomes full
we allocate a new version at the end of $S$ with  twice the size.
We will also allow an operation that creates a new array $A_i$.

\begin{lemma}
\label{SpaceDoublingLemma}
A collection of arrays $A_i[1..n_i], i=1,\ldots, k$
that grows by adding new array entries and creating new arrays 
can be maintained  within an array $S[1..4n]$,
for $n=\sum_i n_i$.
The
extra time is $O(n)$.
\end{lemma}

\begin{proof}
We maintain the invariant that each current 
array $A_i$ 
has size $2s_i$ in $S$ when $n_i\in [s_i+1..2s_i]$, and furthermore
the total amount of space in $S$ used for versions of $A_i$ is
$\le 4s_i$.
(Thus at every point in time every array $A_i$ has  used
$\le 4s_i <4n_i$ entries of $S$, i.e., all allocated storage is within
$S[1..4n]$.)

When a new $A_i$ is created we set $s_i=1$, allocate 2 cells, and set $n_i=2$.

When 
a new entry is added to $A_i$ and
$n_i=2s_i$ we allocate a new copy of $A_i$ of size
$4s_i$ at the end of $S$, copying the contents of $A_i[1..n_i]$ into it.
Setting $s'_i=2s_i$ and $n_i=2s_i+1$
the new $A_i$ has size $4s_i=2s'_i$ with $n_i=s'_i+1$. $A_i$ has used
$\le 4s_i+4s_i=4s'_i$ entries of $S$.
So the invariant is preserved.

The time for the operation (copying $n_i$ entries)
is $O(n_i)$ and can be charged to
the $s_i=n_i/2$ elements added to $A_i$ since the last allocation.
\end{proof}

Returning to Lemma \ref{3.5Lemma} when the final value of $n$ is unknown, note
the space usage consists of
ancestor tables and
single values associated with
each vertex. 
(The children lists for $T$ are stored as vertex values --
$v$ points to its first child, and each child of $v$ points to the next child. The paths of $\cal P$
are also stored as vertex values
$pred(v)$ -- a vertex $v$ on  a heavy path has $pred(v)$ equal to
the predecessor of $v$ on the heavy path.)
The single values are updated in \alm\  operations and recompressions.
Lemma \ref{SpaceDoublingLemma}
 is used to manage all the space (including single vertex values).
We conclude that Lemma  \ref{3.5Lemma} holds in entirety when
$n$ is not known in advance.

Now we extend these algorithms to allow $add\_root$ operations
in addition to \al.. We show that the general incremental tree
problem reduces to  \al. and $ca$ operations.
First extend the characteristic ancestor operation.
For an arbitrary node  $r$ of $T$, let
$nca(x,y;r)$
denote the nearest common ancestor of $x$ and $y$ when
$T$ is rerooted at vertex  $r$. Define $ca(x,y;r)$ similarly.
All other terminology is unchanged, e.g., $ca(x,y)$
 denotes the characteristic ancestors in $T$ with its
original root, and similarly for the term ``ancestor''.
The following lemma shows we can compute $ca(x,y;r)$ just using
the $ca$ functions on $T$.

\begin{lemma}
\label{RootedCalemma}
\i Any 3 nodes in a tree $x,y,z$ have 
$|\{nca(x,y), nca(x,z), nca(y,z)\}|\le 2$. 

\ii
$ca(x,y;z) = \begin{cases}
ca(x,y)       & \text{if }nca(x,z)= nca(y,z),\\
(a, \pt(a), a_y)& \text{if }nca(x,z)=nca(x,y)\ne nca(y,z) \text{ and }
ca(y,z)=(a,a_y, a_z).
\end{cases}$
\end{lemma}

\remark{Part \i and symmetry of $x$ and $y$ show that part
\ii gives a complete definition of $ca(x,y;z)$.}

\begin{proof}
\i Let $a$ be the shallowest of the three nodes $nca(x,y)$, $nca(x,z)$,
$nca(y,z)$. So wlog $a=nca(x,y)$, and let $ca(x,y)=(a,a_x,a_y)$.  If
$a\ne nca(x,z)$ then $nca(x,z)$ is an ancestor of $x$ deeper than $a$.
So $a_x\ne a$ and $z$ descends from $a_x$.
Thus the path from $y$ to $z$ goes through $a$ and $a_x$, so $a=nca(y,z)$.

\ii 
Let $ca(x,z)=(b,b_x,b_z)$.

Suppose $nca(x,z)= nca(y,z)$.  $b=nca(x,z)$ is an ancestor of $y$.  If
$b_x\ne b$ and $b_x$ is an ancestor of $y$ then $nca(y,z)$ descends
from $b_x$, contradicting $b=nca(y,z)$.  Thus $b=nca(x,y)$ and
$b=nca(x,y;z)$. Clearly $ca(x,y;z)=ca(x,y)$.

Next suppose $nca(x,z)=nca(x,y)\ne nca(y,z)$.
The equality implies $b$ is an ancestor of $y$, and the inequality implies
$b\ne b_z$ and $b_z$ is an ancestor of $y$. Thus $nca(x,y;z)=nca(y,z)$.
Let $ca(y,z)=(a,a_y,a_z)$. Then $ca(x,y;z)=(a, \pi(a), a_y)$.
\end{proof}

It is now a simple matter to implement $add\_root$ operations in terms of \al.:
We never change the child lists of the data structure
representing $T$. 
Instead we maintain a pointer $\varrho$ that gives the current root
of $T$ as defined by $add\_root$ operations.
The operation
$add\_root(y)$ is implemented as
\begin{equation}
\label{AddRootEqn}
add\_leaf(\varrho,y);\,\varrho\gets y.
\end{equation}
The algorithm for $ca(x,y)$ is $ca(x,y;\varrho)$.

We close this section by noting that 
$add\_root$ can be implemented directly, without the general reduction.
The main observation is that $add\_root$ changes the compressed tree
in a simple way:
Let $T'$ be the result of performing $add\_root(y)$ on $T$, a tree with
root $x$.
If $|V(T)|>1$ then $y$ plus the heavy path with apex $x$ in $T$
forms a heavy path in $T'$.
Thus $C(T')$ can be constructed from $C(T)$
by changing the name of the root from $x$ to $y$
and giving the root a new child named $x$.
(This works when $|V(T)|=1$ too.)
This transformation is easily implemented in our data structure.

\begin{corollary}
\label{LogSquaredIncNCACor} 
The nearest common ancestors problem with \alm, $add\_root$ and {\em ca}
operations can be solved in $O(m+n\log^2n)$ time and $O(n\log n)$
space.
\end{corollary}

As before the corollary does not require that $n$ be known in advance.

\section{{\em nca}'s for Edmonds' algorithm}
\label{3.2Sec}
This section gives a simple algorithm to find
the nca's needed in Edmonds' matching algorithm.
Each nca operation uses $O(1)$ time,
and 
the total time for all \alm's is
$O(n\log n)$. The extra space is $O(n)$.
So this completes our efficient implementation of the weighted
matching algorithm. (Readers interested 
only in matching needn't go beyond this section.)

This section also introduces the multi-level approach, wherein
the incremental tree algorithm  is  used on a number of 
trees derived from the given tree, 
each at a given ``level''. We use three versions of the approach.
The simplest is for Edmonds' algorithm, and  
the two more elaborate versions are presented in 
 the next two sections.

\begin{figure}[t]
\centering
\input{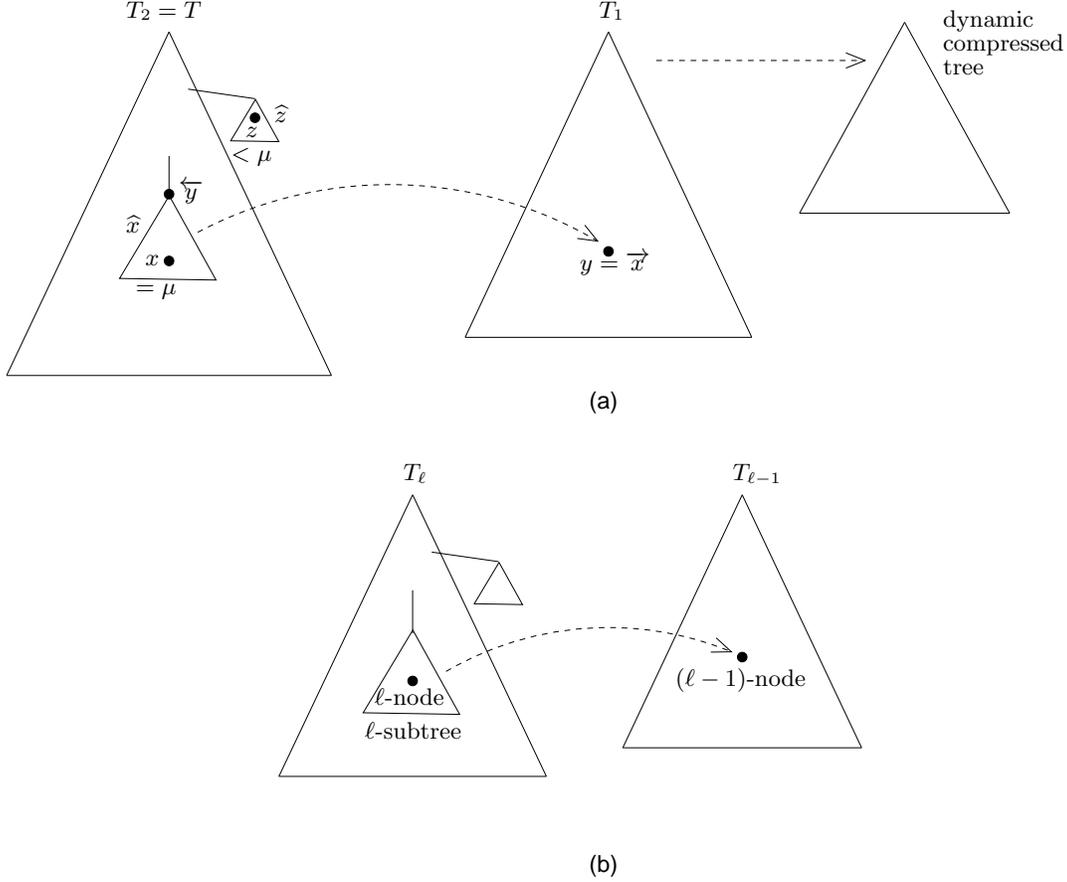}
 \caption{(a) Data structure for $nca$s in
Edmonds' algorithm. $\mu=\f{\log n}$.
$y$ is a 1-node, \wz. is nonfull.
(b) Generalization to $L$ levels,
$T_L=T$, $L\ge \ell >1$.}
 \label{EdLevelsFig}
 \end{figure}

The idea is to reduce the number of tree nodes by contracting small subtrees.
The terms {\it vertex} and {\it the incremental tree} $T$
refer to the objects of the given problem, e.g., an operation
\alm$(x,y)$ makes vertex $x$ the parent of vertex $y$ in $T$.
We  use two trees, illustrated in 
Fig. \ref{EdLevelsFig}(a).
$T_2$ is the incremental tree $T$, enhanced with its data structure.
$T_1$ is a 
smaller version of $T$, derived by contractions and deletions.

This indexing is a special case of our second multi-level structure,
 illustrated in Fig. \ref{EdLevelsFig}(b):
A tree $T$ is represented on $L>1$ levels, with  $T=T_L$, and 
for level $\ell \in [2..L]$
tree $T_{\ell-1}$ a minor of $T_{\ell}$. Edmonds' algorithm uses $L=2$. 

For Edmonds' algorithm define $\mu=\f{\log n}$. 
The algorithm maintains a partition of the vertices of $T=T_2$  
into subtrees of $\le \mu$ vertices called {\em 2-subtrees}.
A 2-subtree
containing exactly $\mu$ vertices is {\em full}.
$T_1$ is formed from $T=T_2$ by discarding the nonfull
2-subtrees and contracting the full ones.
A node of $T_1$ (i.e., a contracted 2-subtree) is called a {\em 1-node}.%

We use this additional notation, illustrated in Fig. \ref{EdLevelsFig}:
For any vertex $x$, \wx. 
denotes the 2-subtree containing $x$.
As an example 2-subtrees are created and maintained in \alm\ as follows:
In \alm$(x,y)$ if \wx. is a full 2-subtree then $y$ is made the root of  new
2-subtree; otherwise $y$ is added to \wx..
(Note this descriptioin guarantees that $T_1$ is a tree and not a forest.)

When \wx. 
is full, \fx. denotes the 1-node containing $x$.%
\footnote{The forward arrow notation corresponds to the
arrows in  Fig. \ref{EdLevelsFig}.}
If $x$ is a 1-node, i.e.~the contraction of a 2-subtree
$S$, \gx. denotes the root vertex 
of $S$ (\gx. is a vertex of $T_2=T$).
Also note that  we write functions of nodes
like
$\pi(x)$ and $ca(x,y)$, relying on context (i.e., the identity of arguments
$x$ and $y$) to indicate 
which tree $T_i$  is being used.

$T_1$ is processed using the incremental-tree $nca$ algorithm of
Lemma \ref{3.5Lemma}. Clearly there are  $O(n/\log n)$ 1-nodes, 
so the time spent on $T_1$ is $O(m+n\log n)$ and the space is $O(n)$.

$T$ uses a simple
data structure: Each root $x$ of a 2-subtree  is marked as such,
and stores the size of its tree $|V(\Px.)|$. 
If $\Px.$ is full then $x$ has a pointer to its  1-node $y=\fx.$;
also $y$ has a pointer to $x=\gy.$.
Each nonroot has a pointer to its 2-subtree root.
Each node $x$ of $T$ has
a parent pointer, as well as 
child pointers;
the children of $x$ that belong to  $\Px.$ 
all occur before the children not in $\Px.$.

Fig.\ref{StaticAlg} gives the algorithm for $nca(x,y)$. 
Note the $ca$ operation takes place in
$T_1$ and the $nca$ operation is in a 2-subtree, in $T_2$.


\begin{figure}
{\parindent=40pt
\def\For{{\bf for }}
\def\KwTo{{\bf to }}
\def\While{{\bf while }}
\def\do{{\bf do }}
\def\If{{\bf if }}
\def\Then{{\bf then }} 
\def\Return{{\bf return }} 
\def\g #1.{\mathy{\overleftarrow{#1}}}

\If $\wx.\ne \wy.$ \Then

{\hi

$/*$ set $x$ and $y$ so $\wx.= \wy.$ and $nca(x,y)$ is unchanged $*/$

\If {\wx. is nonfull} \Then {$x \gets \pi(\rho(\wx.))$};  
 \If {\wy. is nonfull} \Then {$y \gets \pi(\rho(\wy.))$}

$(a,a_x,a_y) \gets ca(\fx.,\fy.)$

\If {$a\ne a_x$} \Then {$x \gets \pi(\g {a_x}.)$}; 
\If {$a\ne a_y$} \Then {$y \gets \pi(\g {a_y}.)$}
}

\Return {$nca(x,y)$}

}

\caption{Finding nca's in Edmonds' algorithm.}
\label{StaticAlg}
\end{figure}

We complete the data structure by showing how to process
\alm\  and $nca$ operations in 2-subtrees. 
We do this by
maintaining 
a representation of ancestors as bitstrings 
in trees that grow by 
\alm\ operations, assuming their size remains $\le \log n$. 
Edmonds' algorithm uses this data structure on every 2-subtree.
The details of the data structure are as follows.

Let $T$ be a tree that grows by \alm\ operations.
The nodes of $T$ are numbered sequentially as they get added, starting at 1.
The number of node $x$ is stored as its
``identifier'' $id[x] \in [1..|V(T)|]$.
$T$ also has an array $v[1..|V(T)|]$
that translates identifiers to their corresponding node, i.e.,
$v[i]$ specifies the vertex of $T$ whose identifier is $i$.

Each vertex $x\in T$ has a RAM word
$anc[x]$ that stores a string of $\le \log n$ bits.
The $i$th bit of $anc[x]$ (corresponding to $2^{i}$) is 1 iff
node number $i$ is an ancestor of $x$. So for example bits 1 and
$id[x]$ are always 1. The key property is that reading the bits 
most-significant-first gives the ancestors of $x$ in their proper order, i.e., decreasing depth.

For \alm\  we maintain a value $s$ as the current size of $T$.
\alm$(x,y)$ is implemented as 
\[\pi(y)\gets x;\;
id[y],\, s\gets s+1;\; v[s]\gets y;\; 
anc[y]\gets anc[x]+2^s.\]
We precompute  a table 
that gives most-significant bits. Specifically
for any bitstring $b\ne 0$ of $\log n$ bits,  
$msb[b]$ is the index of the most significant bit of $b$.
The operation $nca(x,y)$ is implemented as
\[ v\,[msb\,[anc[x]\land anc[y]]].
\]

It is easy to see that \alm\ and $nca$ both use $O(1)$ time.
To use this data structure in Edmond's algorithm,
we keep space usage linear 
by using the doubling strategy 
of Lemma \ref{SpaceDoublingLemma} on the collection of 
$v$ arrays of all  2-subtrees.

Using the results of \cite{G17}, which leaves the incremental-tree
 nca problem as the last detail of Edmonds' algorithm,
we get the following.

\begin{theorem}
\label{EdmondsThm}
A search of Edmonds' algorithm can be implemented in time
$O(m+n\log n)$ and space $O(m)$.
\hfill\qed\end{theorem}



\section{Multi-level incremental-tree algorithms}
\label{3.3Sec}
This section begins by giving the details of the multi-level
approach illustrated in Fig.\ref{EdLevelsFig}(b).
Then it presents a 3-level algorithm to solve the 
incremental-tree nca problem in time $O(1)$ for $nca$ queries,
total time $O(n)$ for \alm\ and $add\_root$ operations,
and
space
$O(n)$. That algorithm is used in the next section
to construct our most general nca algorithm.
It uses the  multi-level structure presented in  this section,
with unbounded number of levels, and
some changes that we note.

\paragraph*{The framework}
This section gives the high-level organization of
an incremental nca algorithm
with an arbitrary number of levels.
We will only need 3 levels in the next section but the number of levels is
unbounded in Section \ref{3.4Sec}.

The terms {\it vertex} and {\it the incremental tree} $T$
refer to the objects of the given problem, e.g., an operation
\alm$(x,y)$ makes vertex $x$ the parent of vertex $y$ in $T$.
A multi-level algorithm works on a number of levels
designated $\ell =L,L-1,\ldots, 1$.

The incremental tree $T$ is 
represented by a tree $T_\ell$ on every level
(a small tree $T$ may have $T_\ell$ empty
for levels $\ell$ less than some threshold).
$T_L$ is  $T$. Every other $T_\ell$ is a smaller tree
derived from $T_{\ell+1}$ by deletions and contractions. 
Each 
$T_\ell$ is composed of {\em $\ell$-nodes} (called {\em nodes} if
the level is clear).
The algorithm maintains a partition of the nodes of $T_\ell$
into subtrees 
called {\em $\ell$-subtrees}. 
Each level is provided with given  algorithms that solve the incremental problem
in any $\ell$-subtree; the multi-level algorithm described in this section 
sews these given algorithms together to solve the incremental problem on the given tree $T$.

Every level $\ell$  has an integral size parameter $\mu_\ell$.%
\footnote{In Section \ref{3.4Sec} these size parameters are replaced
by a notion of ``stage''.}
Every $\ell$-subtree $S$ contains $\le\mu_\ell$ $\ell$-nodes. $S$ is
{\it full} if equality holds.
$\mu_1=n+1$, so $T_1$ is always nonfull (if it exists).
For $L\ge  \ell>1$,
$T_{\ell-1}$ is formed from $T_{\ell}$ by 
discarding every nonfull $\ell$-subtree and
contracting the full ones.
The fact that any $T_\ell$ is  tree (i.e., not a forest) follows from
this invariant: 
For every level $\ell$,
the nonfull $\ell$-subtrees of $T_\ell$ are at its frontier, i.e.,
any node $x$ in a nonfull $\ell$-subtree $S$ 
has all its $T_\ell$-children in $S$.

Efficiency of a multilevel algorithm is
 achieved using the shrinkage of the tree from level to level.
Specifically an $\ell$-node 
with $\ell<L$ contains 
$\Pi_{\ell+1}^L \mu_i$ 
vertices of $T$.
So the number of $\ell$-nodes is
\begin{equation}
\label{NumEllNodes}
|V(T_\ell)|\le 
n/\Pi_{\ell+1}^L \mu_i 
.
\end{equation}

We use this additional notation:
Let $x$ be an $\ell$-node, $L\ge \ell\ge1$.
\Px. denotes the $\ell$-subtree containing $x$.  
If $\Px.$ is full (in particular
$\ell>1$) then $\fx.$ denotes the ($\ell-1$)-node that is the
contraction of $\Px.$.  If $\ell<L$ then $x$ is the contraction of an
$(\ell+1)$-subtree $S$, and $\gx.$ denotes the root node of $S$. 
 As before we write functions of nodes like $\pi(x)$,
relying on context (i.e., the identity of argument $x$) to indicate
which tree $T_\ell$ is referenced.

Each $T_\ell$ uses this data structure: 
Let $x$ be an  $\ell$-node. 
If $x$ is the root of its $\ell$-subtree
it stores the subtree size $|V(\Px.)|$ 
($1\le |V(\Px.)|\le \mu_\ell$). It also
has pointers to $\gx.$ if $\ell<L$ and
 $\fx.$ if $\ell>1$.
A nonroot $x$ has a pointer to its $\ell$-subtree root.
Every $x$ has a parent pointer.

The main routines for incremental trees are
\[a(x,y,\ell),\ \wa(x,y,\ell),\ c(x,y,\ell),\ \wca(x,y,\ell).
\]
$a(x,y,\ell)$ is a recursive routine that performs the \alm\ operation
for $\ell$-node $x$ and new $\ell$-node $y$. 
It  may call $a(\fx.,\fy.,\ell-1)$.
For vertices $x,y$ in the given graph the operation
$\alm(x,y)$ is performed by $a(x,y,L)$. 
The $a$ routine makes use of $\wa$ to grow $\ell$-subtrees. Specifically
for $\wa(x,y,\ell)$, $x$ is an $\ell$-node
and
$y$ a new $\ell$-node that is made a child of $x$ in the $\ell$-subtree
$\Px.$.

The  $ca$ operation is organized similarly. It uses
the recursive routine $c(x,y,\ell)$, which 
returns the characteristic ancestors of $\ell$-nodes 
$x$ and $y$, possibly invoking $c(\fx.,\fy.,\ell-1)$.
For vertices $x,y$ in the given graph the operation
$nca(x,y)$ is performed by $c(x,y,L)$. 
The $c$ routine makes use of
$\wca$, which returns the characteristic ancestors of $\ell$-nodes 
$x$ and $y$ that belong to the same $\ell$-subtree
$\wx.=\wy.$.

We will extend these operations below to allow $add\_root$.
Also looking ahead, 
in Section \ref{3.4Sec}
the $nca$ routine will  use $c$ and  $\wca$
with some obvious modifications.
The {\em link} routine will
use the same overall structure 
as $a$ and $\wa$
for \alm.

Now we describe the two recursive algorithms
starting with $a(x,y,\ell)$:

{\narrower

{\parindent=0pt

\case {$\Px.$ is full}
Make $x$ the parent of node $y$,
and make $y$ a singleton $\ell$-subtree.

\case {$\Px.$ is nonfull} Execute $\wa(x,y,\ell)$.
If $\Px.$ is still not full we are done but suppose
$\Px.$ has become full.
Create a new 
$(\ell-1)$-node $z$. Make $z$ the node $\fx.$.
Now
 there are two subcases:

\subcase {$\Px.=T_\ell$}
Make $z$ the unique $(\ell-1)$-node, as well as
a singleton $(\ell-1)$-subtree.

\subcase {$\Px.\ne T_\ell$}
$w=\pt(\rt(\Px.))$ is in a full $\ell$-subtree.
Execute
$a(\fz w.,z,\ell-1)$ to 
add $z$ as a new $(\ell-1)$-leaf.

}}

\b

The \alm\ algorithm preserves the defining properties of $T_\ell$
trees and so is correct.  The total time for all \alm\ operations is
dominated by the time used by $\wa$ to build all the $\ell$-subtrees,
$L\ge \ell\ge 1$. (This includes the time to create a new
singleton subtree.)


We turn to the $c$ routine.
The high-level strategy is simple --
use  $c(\fx.,\fy.,\ell-1)$ 
to find the $\ell$-subtree  $S$ containing $nca(x,y)$,
and then use the
 $\wca$ routine in $S$ 
to find the desired level $\ell$ characteristic ancestors
$(c,c_x,c_y)$.
There are various special cases, depending on whether or not
\wx. and \wy. are full, whether or not
$S$ contains $c_x$ or $c_y$, etc.
The details of  $c(x,y,\ell)$ 
are as follows.

{\narrower

{\parindent=0pt

\case {$\Px.$ and $\Py.$ are both full}
Fig.\ref{MultiAlg} gives pseudocode for this case. It handles
special cases such as $x=y$ or $\wx.=\wy.$.


\begin{figure}

{
\def\For{{\bf for }}
\def\KwTo{{\bf to }}
\def\While{{\bf while }}
\def\do{{\bf do }}
\def\If{{\bf if }}
\def\Then{{\bf then }} 
\def\Return{{\bf return }} 

\def\g #1.{\mathy{\overleftarrow{#1}}}

\def\myif #1 #2 {{\bf if} #1 {\bf then} #2}
\def\myifel #1 #2 #3 {{\bf if} #1 {\bf then} #2 {\bf else} #3}

\parindent=40pt

$(a,a_x,a_y) \gets c(\fx.,\fy.,\ell-1)$

\myif {$a_x\ne a$} {$x \gets \pi(\gz {a_x}.)$; }
 \myif {$a_y\ne a$} {$y \gets \pi(\gz {a_y}.)$}

{$(b,b_x,b_y)\gets \wca(x,y,\ell)$}

\myif {$b_x=b$ and $a_x\ne a$} {$b_x\gets \gz {a_x}.$; } 
 \myif {$b_y=b$ and $a_y\ne a$} {$b_y\gets \gz {a_y}.$}

\Return {$(b,b_x,b_y)$}

}

\caption{Procedure for $c(x,y,\ell)$ when $\wx.$ and $\wy.$ are full.}
\label{MultiAlg}
\end{figure}

\case
{One or both of $\Px.,\Py.$ is nonfull}
If $\wx.=\wy.$ we use 
$\wca(x,y,\ell)$ directly. (This includes the special case where
there are no $(\ell-1)$-nodes.)
Assuming $\Px.\ne \Py.$, 
when $\wx.$ is nonfull 
 we replace $x$ by
$\pt(\rt(\Px.))$, 
and similarly for $y$. We then  execute
the code of
Fig.\ref{MultiAlg}.
If the returned $b_x$
is the replacement for $x$ we change $b_x$ to 
$\rt(\Px.)$, and similarly for $y$.

}}

\b

The analysis of
this algorithm is similar to \alm:
Correctness follows from the defining properties of
$T_\ell$ trees.
The time for an 
operation $nca(x,y)$ is
dominated by the time used by the routines
$\wca(x,y,\ell)$,
$L\ge \ell\ge 1$.

We extend  the routines to allow $add\_root$ 
similar to the extension for Corollary
\ref{LogSquaredIncNCACor}, 
as follows. $add\_root$ 
is still implemented by \eqref{AddRootEqn}, where now
 $\varrho$ is a pointer to a node in the tree $T_L$.
The routine for $nca(x,y)$, instead of immediately calling $c(x,y,L)$,
is modified to use
Lemma \ref{RootedCalemma}\ii as before. Specifically 
it calls
$c(x,y,L),c(x,\varrho,L)$, and $c(y,\varrho,L)$,
and chooses $nca(x,y)$ according to the lemma.

\paragraph*{Linear-time incremental trees}
Take $L=3$ levels with
\begin{equation}
\label{muDefEqn}
\mu_3= \mu_2= \c{\log n},\ \mu_1=n+1.
\end{equation}
Level 1 uses the incremental-tree algorithm of Section \ref{3.1Sec}.
It uses $O(m+n)$ time and $O(n)$ space. This follows since
\eqref{NumEllNodes} shows there are $\le {n\over \log^2 n }$ 1-nodes.
Thus 
Corollary \ref{LogSquaredIncNCACor} 
shows the 
time on level 1 is $O(m+{ n\log^2 n \over  \log^2 n  } )=O(m+n)$. 
The space in level $1$ is 
$O( { n\log n \over \log^2 n  } )=O(n)$.

Levels 3 and 2 both use the bitstring data structure of the previous section.
For level 2  we must extend the $nca$ algorithm to compute all characteristic
ancestors $ca$.
We precompute  a table 
that gives least-significant bits. Specifically
for any bitstring $b\ne 0$ of $\log n$ bits,  
$lsb[b]$ is the index of the least significant bit of $b$.
The operation $ca(x,y)$ is  implemented as
\[
{\bf if\ } nca(x,y)=x {\bf\ then\ } c_x=x {\bf \ else\ } c_x =
 v\,[lsb\,(anc[x]\land \neg anc[y])].
\]

This discussion shows that
levels 3 and 2 use $O(m+n)$ time.
The space is kept linear, $O(n)$,
by using the doubling strategy 
of Lemma \ref{SpaceDoublingLemma}
for
the $v$ tables of nonfull $\ell$-subtrees.

This completes the
3-level incremental tree algorithm.

\begin{theorem}
\label{3.1Thm} 
The incremental-tree nearest common ancestors problem
with 
\alm, $add\_root$ and
$ca$ operations
can be solved in  $O(m+n)$ time and  $O(n)$ space.\hfill$\Box$
\end{theorem}

As in Corollary \ref{LogSquaredIncNCACor}  the theorem does not require $n$ (the number of
\alm\  and $add\_root$ operations) to be known in advance.
To achieve this first consider 
\eqref{muDefEqn} defining
the 
$\mu_i$. One approach is to update these values every time $n$ doubles.
Instead we will simply interpret $n$ in
\eqref{muDefEqn} to be
$N$,
the maximum integer that can be represented in the RAM. So
$\mu_3=\mu_2=\log N$ is the number of bits in a RAM word. 
The timing estimates are unchanged. For instance the time in level 1
is
$O(m+{ n\log^2 n \over  \log^2 N  } )=O(m+n)$. 
The space in level $1$ is 
$O( { n\log n \over \log^2 N  } )=O(n)$.
The space for all three levels is maintained in one array $S$,
using Lemma \ref{SpaceDoublingLemma}.


\section{Link operations}
\label{3.4Sec}
This section extends the multi-level data structure to solve our most general
dynamic nca problem.
The algorithm processes $m$ $nca$ and {\it link} operations
on a set of $n$ nodes in time $O(m\alpha(m,n)+n)$
and linear space $O(n)$.

The multilevel structure shares many details with that of the previous section:
The levels $\ell=L,\ldots, 1$, the notions of $\ell$-tree,
$\ell$-node, and $\ell$-subtree are all unchanged.
A difference is that a tree $T_\ell$ at level $\ell>1$ gives its
 level $\ell-1$ counterpart $T_{\ell-1}$ by contracting {\em every} 
$\ell$-subtree, i.e., no subtrees are deleted.
The notations $\Px., \fx.$, and $\gx.$ are defined without change.

$nca$ operations are
implemented using 
the $c$ and $\wca$ routines as in last section. 
$link$ operations are
implemented
using a recursive routine $l$ similar to $a$ of last section.
The analog of \wa\ for link is folded into $l$, i.e., there is no
$\wl$. It is convenient to
use an extra argument for $l$: We write $l(r,x,y,\ell)$ where $r$ is the root
of the $\ell$-tree containing $x$.
Call a tree built up by {\it link} operations a {\em link tree}.
The operation $link(x,y)$ is performed by $l(\rho, x,y,L)$ for $\rho$ the root
of the link tree containing $x$.

For motivation we 
start by sketching the two simplest 
versions of our algorithm.

\long\def\example #1. #2{\bigskip \noindent{\bf Example: #1.}
{#2}\bigskip}

\example {Algorithm 1}. {Every link tree is represented as an incremental tree.
The {\em link} operation uses
{\em add\_leaf}  and {\em add\_root} operations to 
transfer the nodes
of the smaller tree into the larger (for trees of equal size
break the tie arbitrarily). It then  discards the
incremental tree of the smaller tree. 
The number of node transfers is $O(n\log n)$.
So Theorem \ref{3.1Thm} shows the 
total time is $O(m+n\log n)$.}

\def\stage.{\mathy{\sigma}}

\def\ua{\hskip-2pt\uparrow\hskip-2pt}

The analysis of Algorithm 1 is based on what we will call the ``stage''
of the link tree: A tree in stage \stage. has 
between $2^{\stage.}$ and 
$2^{\stage.+1}$ vertices. We view the analysis
as charging a vertex $O(1)$ to advance from one stage to the next. 
(This accounts for the total time spent on \alm\  and $add\_root$ operations,
since Theorem \ref{3.1Thm} shows the time spent on a tree that ultimately
grows to $n_s$ nodes is $O(n_s)$, i.e., the time is proportional to the number
of node transfers.)
Our more efficient
algorithms maintain explicit stages, and these stages
will require faster growth in the tree size.

\example {Algorithm 2}.
{Algorithm 1 can be improved using a 
2-level stategy similar to previous ones.
Level 2 classifies each tree as stage 1 or 2: A 
2-tree  is in  stage 1 
if it has $<\log n$ nodes and stage 2 if it has
$\ge \log n$ nodes.

A stage 2 2-tree is partitioned into 2-subtrees, each of which contains
$\ge \log n$ nodes.
Each 2-subtree is represented as an incremental tree, 
using the data structure of Theorem \ref{3.1Thm}.
Contracting all these 2-subtrees gives its corresponding 1-tree. 

A stage 1 2-tree is also a 1-tree. For consistent terminology in stage 1,
view each 2-node as a 2-subtree.

Level 1  uses Algorithm 1 on all 1-trees. 

The $l$ routine works as follows on level 2:
It sets $\pi(y)\gets x$. Then, letting $X$ and $Y$ denote the
2-trees containing $x$ and $y$ respectively, it
executes the case below that applies:

\case {Both trees are in stage 2} Link the level 1 trees using Algorithm 1.
 
\case {Only one tree is stage 2} If $X$ is stage 2,
transfer the nodes of $Y$ to \wx.,
using \alm\ operations. The discard the data structures for
$Y$ on levels 1 and 2. 

If $Y$ is stage 2 do the same, using appropriate
$add\_root$ operations in the transfer of $X$ to $\wh y$. 

\case {Both trees are stage 1} 
If the combined trees contain
$\ge \log n$ nodes initialize the 2-tree  as a 
new stage 2 tree, 
with one 2-subtree consisting of all nodes of $X$ and $Y$. 
Discard the data structures for $X$ and $Y$ on both levels.

Otherwise link the level 1 trees using Algorithm 1.

\b

The total time is dominated by the time spent for all incremental trees on
both levels 1 and 2.
On level 2  a 2-subtree that grows to contain $n_i$ nodes
(as in the last two cases) uses time $O(n_i)$ for all
\alm\ and $add\_root$ operations.
So all 2-subtrees use total time $O(n)$.

Consider  level 1.
The 
1-trees for
stage 2 2-trees
collectively contain $\le n/\log n$ nodes.
Each node is transferred by Algorithm 1
at most $\log n$ times. So the total
time is $O(n)$.
The stage 1 2-trees collectively contain $n$ nodes. Each is transferred
$\le \log \log n$ times by Algorithm 1. 
So the total time is $O(n\log\log n)$.
This term strictly dominates the algorithm's time bound.

Clearly we can improve this algorithm by adding another stage, for
2-trees with $\le \log \log n$ nodes. The time becomes 
$O(n \log^{(3)} n)$. 
Continuing in this fashion we can achieve time  $O(n\log ^* n)$.%
\footnote{$\log^{(i)} n$ and $\log ^* n$ are defined as in \cite{CLRS}.} 
Let us sketch this algorithm.
(The detailed version of the algorithm is the case $\ell=2$ of
algorithm ${\cal A}_\ell$ presented below.)
It is convenient to switch notation from 
small functions like $\log$ to large ones like exponentiation.
Recall the superexponentiation function, defined by
$2\ua 1 = 2$, $2\ua(s+1) = 2^{(2\uparrow s)}$.

In Algorithm 2 level 2 has $\log^* n$ stages.  A stage \stage. 2-tree
has between $2\ua \stage.$ and $2\ua (\stage.+1)$ nodes.  It
is partitioned into 2-subtrees, each of which
contains $\ge 2\ua\stage.$ nodes.  The remaining properties of 2-trees
are essentially the same as the previous algorithm.

The $l$ routine uses new criteria to determine the cases, but is
otherwise unchanged.  
In more detail, let $X$ be in stage $\sigma(X)$ and similarly
for $\sigma(Y)$, and let $\sigma=\max\{\sigma(X),\sigma(Y)\}$.  

If the
combined trees contain $\ge 2\ua(\sigma+1)$ nodes a new stage
$\sigma+1$ tree is initialized (as in the last case above).

Otherwise
if $\sigma(X)\ne \sigma(Y)$ the nodes of the smaller 2-tree are
transferred to the larger (as in the middle case above).  

Otherwise
($\sigma(X)= \sigma(Y)$) Algorithm 1 links the images of the two
2-trees (as in the first and last cases).

The time for all $link$ operations is $O(n\log^*n)$. This holds
because the time on each stage is $O(n)$. Let us sketch a proof.
Consider  level 2. As before, a  2-subtree that grows to contain $n_i$ nodes
uses time $O(n_i)$ for all
\alm\ and $add\_root$ operations.
This gives $O(n)$ time total for each stage on level 2. 
There are $\log^*n$ stages so the total time is $O(n\log^* n)$.

As for level 1, a 1-node is a contracted 2-subtree.
A fixed stage $\stage.$ of level 2
contains a total of $\le n/(2\ua \stage.)$ 2-subtrees.
Thus over the entire algorithm stage $\stage.$ has $\le n/(2\ua \stage.)$ 
1-nodes $x$.
After being transferred $2\ua \stage.$ times by Algorithm 1, 
$x$'s 1-tree has
grown to $\ge 2^{ 2\uparrow \stage.}=2\ua (\stage.+1)$ 1-nodes. 
So the 2-tree containing $\gx.$
has advanced to stage $\stage.+1$.
So $O(2\ua \stage.)$ time total is spent on $x$ in level 1.
Thus the time for Algorithm 1 to process all stage $\stage.$ 1-nodes is
$O( \frac{n}{2 \ua \stage.}\,\cdot\, {  2\ua \stage. } )=O(n)$. 
Again there are $\log^*n$ stages so the total time 
on level 1 is $O(n\log^* n)$.

We conclude that
Algorithm 2 uses total time
$O(m+n\log^* n)$.
} 

This construction can be repeated, using Algorithm 2 to get even faster
Algorithm 3, etc. We now present the formal details.
Define Ackermann's function $A_i(j)$ for $i,j\ge 1$ by
\begin{eqnarray*}
A_1(j)&= &2^j, \hbox{\ for\ } j \ge 1;\\
A_i(1)&=&2, \hbox{\ for\ } i \ge 2;\\
A_i(j) &= &A_{i-1}(A_i(j-1) ), \hbox{\ for\ } i,j \ge 2.
\end{eqnarray*}
\noindent Define two inverse functions,
\begin{eqnarray*}
a_i(n)&=&\min\set {j}{A_i(j)\ge n};\\
\alpha(m,n)&=&\min\set{i}{A_i(4\c{m/n})\ge n}, 
\hbox{\ for\ } m,n \ge 1.
\end{eqnarray*}
These definitions differ slightly 
from those of \cite{T83} but this does not change asymptotic estimates. The
most significant difference is that our function $A_i(1)$ is constant
compared to a rapidly growing function in \cite{T83}. This makes for a more
convenient treatment of the base case in our algorithms.
We use some very weak properties of Ackermann's function including these
inequalities, which are proved in Appendix \ref{AckAppendix}:
\begin{eqnarray}
\label{4Eqn}
A_i(j+1) &\ge &2A_i(j), \hbox{\ for\ } i,j \ge 1;\\
\label{5Eqn}
A_{i+1}(j) &\ge &A_i(2j), \hbox{\ for\ } i \ge1,j \ge 4;\\
\label{6Eqn}
\alpha(m',n')&\ge&\alpha(m,n)-1, 
 \hbox{\ for\ } m'\le 2m, n'\ge n.
\end{eqnarray}

A preprocessing step  tabulates the relevant values of
Ackermann's function. We use the values
$A_i(j)$ that are $\le n$ for $i\le \alpha(m,n)$.
Define an array
$ackermann[1..\log n,1..\log n]$: If
$A_i(j)\le n$ then $ackermann[i,j]= A_i(j)$, else $ackermann[i,j]= \nil$.
This table also allows us to find $\alpha(m,n)$,
which is $\le \log n$ (since
\eqref{5Eqn} shows $A_{\log n}(4)\ge A_1(2^{1+\log n})$).
The table is initialized, and  $\alpha(m,n)$ is found, in time
$O(\log^2 n)$. The table allows
any desired value of Ackermann's function to be found in
$O(1)$ time.

We use the linear-time
incremental tree data structure of Theorem \ref{3.1Thm}. Call a tree that is
represented by this data structure an {\it incremental tree}.
The preprocessing step computes all the tables for this algorithm
in time $O(n)$.


The approach is similar to that of \cite{G85b} for a list-splitting problem.
We construct a family of algorithms 
${\cal A}_\ell$, $\ell\ge 1$.
${\cal A}_\ell$
is a multi-level algorithm based on the function $A_\ell$. It
calls ${\cal A}_{\ell-1}$ if $\ell> 1$. 
${\cal A}_\ell$ runs in time $O( m\ell+na_\ell(n) )$.

Algorithm ${\cal A}_\ell$ works on level $\ell$.
The terms {\it $\ell$-node} and {\it $\ell$-tree}
refer to the objects manipulated by ${\cal A}_\ell$.
Every link tree corresponds to an $L$-tree with the same nodes and
edges.
Every level $\ell$ 
has $a_\ell(n)$ {\it stages} $\sigma$, $\sigma=0,\ldots, a_\ell(n)-1$.
Each $\ell$-tree $T$ belongs to  a unique stage $\sigma$ defined as follows.


\case {$|V(T)|<4$} $T$ is in stage  $\sigma=0$.
Stage 0 uses a trivial algorithm so
an invocation of $c$ or $\ell$ uses time $O(1)$.

\case {$|V(T)|\ge 4$} $T$ is in the stage $\sigma\ge 1$ satisfying
$|V(T)|\in [2A_\ell(\sigma),2A_\ell(\sigma+1))$.
(This is possible since $A_\ell( 1)=2$.)
An {\it $\ell$-subtree in stage $\sigma$}
is a subtree that has $\ge 2A_\ell(\sigma)$ nodes. 
The nodes of $T$ are partitioned into $\ell$-subtrees.
If $\ell>1$ then $T$, with each 
$\ell$-subtree contracted, is represented on level
$\ell-1$.

Note that the contracted tree on level $\ell-1$ may be a trivial tree
in stage 0 (of level $\ell-1$). Also if
$\ell=1$ there is no need to store the contracted tree
since $T$  has only one $\ell$-subtree. This follows  since
an $\ell$-subtree has $\ge 2A_1(\sigma)=2^{\sigma+1}$ nodes and
$|V(T)|<2A_1(\sigma+1)=2^{\sigma+2}$ nodes.

\b

Algorithm ${\cal A}_\ell$ uses the following data structure.
Each $\ell$-tree and $\ell$-subtree is represented by its root.
An $\ell$-tree $T$ is stored using parent pointers and children lists.
If $r$ is the root of $T$ then $s(r)$ equals the size of $T$ (the number of its
$\ell$-nodes).
For any node $x$, $\sigma(x)$ equals the stage
of $x$'s $\ell$-tree; if $\sigma(x)>0$ then
$\wh x$ points to the $\ell$-subtree containing $x$.
Each $\ell$-subtree is represented as an incremental tree.
Recall  (Theorem \ref{3.1Thm}) that 
it has a root pointer $\varrho$
which is updated by $add\_root$ operations.

We turn to the 
$link$ and $nca$ operations.
Initially every node is a singleton link tree, in stage 0 of level $L$.
Recall the operation
$link(x,y)$ is processed by 
invoking 
the recusive algorithm 
$l(r,x,y,\ell)$ with arguments 
$\ell=L$ and $r$
equal to
the root of the link tree containing $x$.
$r$ is found by a simple
recursive algorithm: 
If $\rho(\wx.)$ is the root of its $\ell$-tree then it is $r$.
Otherwise
recursively 
compute $r$ as the root of the $(\ell-1)$-tree containing $\fx.$ 
and set $r\gets \gz r.$.

The algorithm for $l(r,x,y,\ell)$ is as follows.
Let $X$ and $Y$ denote the $\ell$-trees with root $r$ and $y$ respectively,
on entry to $l$.

\b

\noindent
{\em Combine Step}:
Combine $X$ and $Y$ to a new $\ell$-tree $T_\ell$ by
setting $\pi(y)\gets x$ and adding $y$ to the child list of $x$.

\b

The rest of the algorithm determines the stage of $T_\ell$ and its decomposition
into $\ell$-subtrees.
Start by
increasing $s(r)$ by $s(y)$. Let $\sigma=\max\{\sigma(x), \sigma(y)\}$.
Execute the first of the  following cases that applies
and then return.

{

\numcase 1 {$s(r)\ge 2A_\ell(\sigma+1)$} 
Make $T_\ell$ a new stage $\sigma+1$ $\ell$-tree consisting of one $\ell$-subtree, as follows:
Initialize a new incremental tree $\wh r$.
Traverse $T_\ell$ top-down; when visiting a node $v$ do an \al. operation
to add $v$ to $\wh r$.
Discard the data structures for $X$ and $Y$
on all levels $\le \ell$. 
If $\ell >1$ then
create an $(\ell-1)$-tree in stage 0 for $T_\ell$, consisting of
one node.

\numcase 2 {$\sigma(x)>\sigma(y)$} 
Traverse $Y$ top-down, doing \al. operations to add each node of $Y$
to the incremental tree $\wh x$.
Discard the data structures for $Y$ on all levels $\le \ell$.

\numcase 3 {$\sigma(x)<\sigma(y)$}
Traverse the path from $x$ to $r$ in $X$, doing $add\_root$ operations to
add each node to the incremental tree $\wh y$.
Then traverse $X$ top-down,
doing \al. operations to add the other nodes to $\wh y$.
Discard the data structures for $X$ 
on all levels $\le \ell$ and set $\sigma(r)\gets\sigma(y)$.

\numcase 4 {$\sigma(x)=\sigma(y)$} 
If $\sigma>0$ then do $l(\fz r., \fx., \fy.,\ell-1)$.
If $\sigma=0$ then combine $X$ and $Y$ using a trivial algorithm.

}

\begin{figure}[t]
\centering
\input{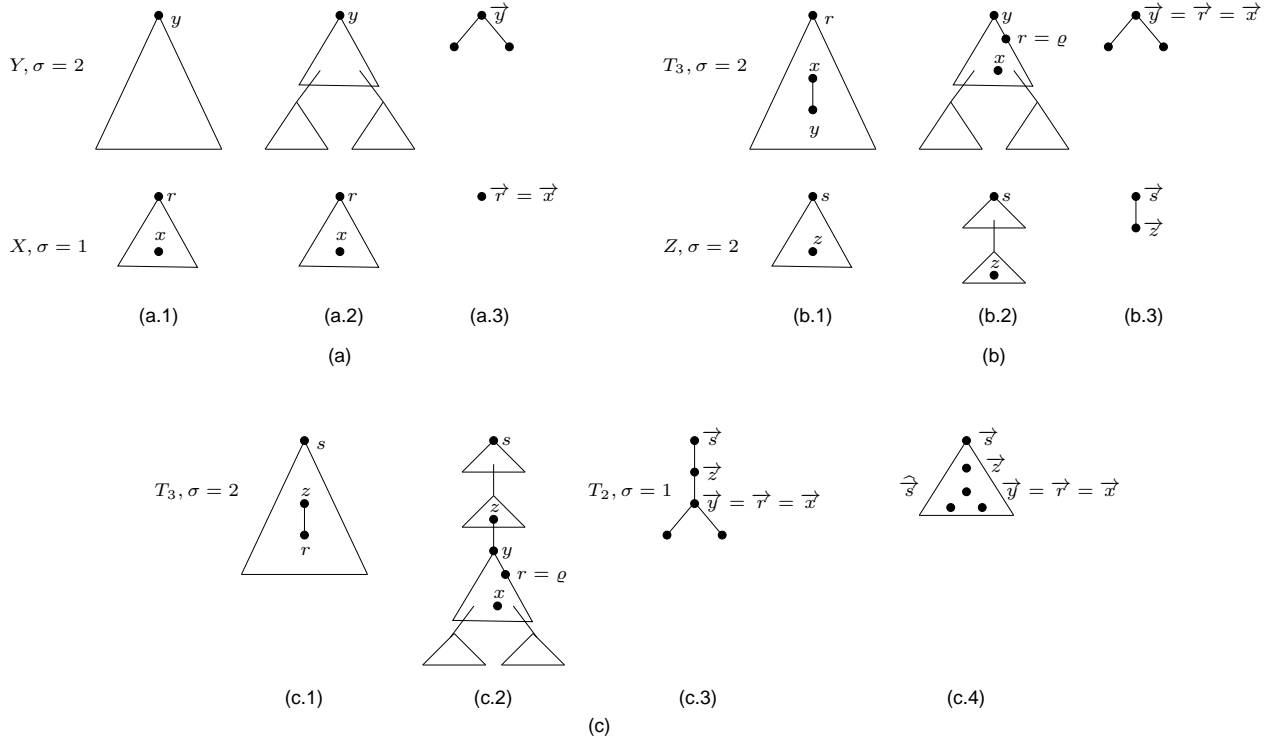}
 \caption{Examples for $link$. $L=3$. (a) Trees for $link(x,y)$:
(a.1) 3-trees $X$ and $Y$. (a.2) 3-subtrees for $X$ and $Y$. $X$ has one
3-subtree. (a.3) 2-trees
for $X$ and $Y$. (b) $T_3$ formed for  $link (x,y)$, and link tree $Z$.
(b.1)--(b.3) 3-trees, 3-subtrees, and 2-trees, as before.
(c) $T_3$ formed for $link(z,r)$. (c.1)--(c.3) 3-tree, 3-subtrees, 2-tree.
(c.4) 2-subtree contains entire 2-tree.}
 \label{AlphaFig}
 \end{figure}

\b

Fig.\ref{AlphaFig} illustrates two $link$ operations.
$link(x,y)$ starts with the trees of Fig.\ref{AlphaFig}(a)
and executes Case 3 in level 3.
The resulting link tree
$link(x,y)$ has root $r$, which is the nonroot node
$\varrho$ in the incremental tree of Fig.\ref{AlphaFig}(b.2).
$link(z,r)$ starts with the trees of Fig.\ref{AlphaFig}(b)
and executes Case 4 in level 3. Then it executes Case 1 for the two stage 0
trees in level 2.
The resulting 2-tree (Fig.\ref{AlphaFig}(c.3)) is in stage 1 since it has $5\ge 4$ nodes.
All 2-nodes are in one 2-subtree, as in Fig.\ref{AlphaFig}(c.4).
 Fig.\ref{AlphaFig}(b.2) and (c.2) illustrate that in general, any
incremental tree may have its root pointer $\varrho$ pointing to
a node at arbitrary depth.

\begin{lemma}
The algorithm for $link(x,y)$ preserves all 
 the defining properties of the data structure. 
\end{lemma}

\begin{proof}
We sketch the argument giving only the most interesting details.
Assume the bookkeeping fields $\sigma(v)$ and \wv.
are updated when node $v$ is added to a new incremental tree.

Consider a link tree $T$. For every level $\ell\in [1..L]$
$T_\ell$ denotes the corresponding tree as defined by
the data structure's parent and child pointers.  
$T$ is represented correctly by  $T_L$, by a simple induction using
the Combine Step.
Furthermore for every level $\ell \in [2..L]$ $T_{\ell-1}$
is formed from $T_{\ell}$ by contracting every $\ell$-subtree.
(Notice that in Case 2 $T_{\ell-1}$ does not change since
$Y$ is absorbed into $\Px.$. Similarly for Case 3 and $\Py.$.)
This implies that the vertex $y$ is the root of its tree $T_L$,
and $\fy.$ and all lower images are roots of their $\ell$-trees
(see especially Fig.\ref{AlphaFig}(b.2)).
This justifies the argument $\fy.$ in the recursive call of Case 4.

Now consider the four cases that determine the stage and the $\ell$-subtrees.

\case {\rm 1} The new $\ell$-tree belongs in  stage  $\sigma+1$
because $s(r)< 4A_\ell(\sigma+1)\le 2A_\ell(\sigma +2)$ by 
\eqref{4Eqn}.

\case {\rm 2} The incremental tree $\wh x$ exists, since 
$x$ is in a positive stage. Similarly in Case 3 $\wh y$ exists.

\case {\rm 4} The new $\ell$-tree $T_\ell$
(formed in the Combine Step) is correctly  partitioned into 
$\ell$-subtrees, since $X$ and $Y$ were. Also 
Case 4 always has $\ell>1$. 
(So level ${\ell-1}$ actually exists.) 
This is because if $\ell=1$ and $\sigma(r)=\sigma(y)=\sigma$ then 
Case 1 applies, since
$s(r)\ge 2(2A_1(\sigma))=
2^{\sigma+2}=2A_1(\sigma+1)$. 

If $\sigma=0$ then $s(r)<4$ so having updated $T_\ell$ to the combined
tree  we are done.
\end{proof}

The algorithm for $ca(x,y)$ is trivial in universe zero.
In positive universes it is the multi-level algorithm 
$c(x,y,\ell)$ of Section \ref{3.3Sec}.
The first case of the $c$ algorithm
is always used
(since every $\ell$-subtree is contracted to 
an $(\ell - 1)$-node).
It executes the code of Fig.\ref{MultiAlg}.

\begin{lemma}
\label{3.6Lemma} 
Algorithm ${\cal A}_\ell$ executes a sequence of $m$ $ca$
and {\it link} operations on a set of $n$ nodes in $O(m\ell+na_\ell(n))$ time
and $O(na_\ell(n))$ space.
\end{lemma}

\begin{proof}
First consider the time.
A $ca$ query uses $O(\ell)$ time in a positive stage, since
$O(1)$ time is spent on each of $\ell$ levels of recursion.
The time is $O(1)$ in universe zero.

The time for {\it links} is estimated as follows.
Charge each {\it link} operation $O(\ell)$ time to account for the 
initial computation of the root $r$ plus the $\ell$ levels of recursion 
and associated processing in routine $l$ (e.g., Case 4 and the Combine Step).
So far all charges are included in the term $O(m\ell)$ of the lemma.

For the rest of the time call an \al. or $add\_root$ operation an $add$ operation and define
\[
\eta = \text{the total number of $add$ operations.} 
\]
($\eta$ includes all $add$ operations done in recursive calls.)
The rest of the time for $l$ is proportional to $\eta$. 
Here we are using
Theorem \ref{3.1Thm}, which  shows that
an incremental tree that grows to contain $n_i$ nodes
uses time $O(n_i)$ for all \alm\ and $add\_root$ operations.
(Also note that discarding data structures in Cases 1--3 is just a nop.)
For the time bound of the lemma  it suffices to show  $\eta=O(na_\ell(n))$.
In fact we will show by induction on $\ell$
that 
\begin{equation}
\label{EtaEqn}
\eta\le 2 na_\ell(n).
\end{equation}

First consider the $add$ operations 
in Cases 1--3 of level $\ell$, i.e., we
exclude the operations that result from a recursive call 
made in Case 4 from level $\ell$. 
Each such $add$ is done for a node
previously in a lower stage of level $\ell$.
So at most one $add$  is done for each node in each stage.
This gives $\le na_\ell(n)$ $add$s total.
(In particular this establishes the base case of the induction,
$\ell=1$.)

To bound the number of $add$s in all levels $<\ell$,
fix a stage $\sigma>0$ of level $\ell$. 
We will show there are $\le n$ $add$s
total in recursive calls made from  stage $\sigma$ of level $\ell$.
So the 
$a_\ell(n)$ stages contribute a total of
$\le na_\ell(n)$ $add$s in levels $<\ell$.
Adding together the two bounds gives
\eqref{EtaEqn} and
completes the induction.

First note an approach that does {\em not} work.
The inductive assumption holds for $\A._{\ell-1}$.
So we could estimate the total number of 
$(\ell-1)$-nodes,
say $n_{\ell-1}$,
and use the inductive bound $2  n_{\ell-1} a_{\ell-1}(n_{\ell-1})$.
But this
overestimates the number $add$s, since 
$\A._\ell$ discards the entire data structure for an $\ell$-tree
as soon as it moves to a higher stage, i.e., when it has size
$\ge 2A_\ell(\sigma+1)$ in Case 1, or earlier in Cases 2 and 3. 
So instead, our approach is to  count the number of $add$s for
each maximal $\ell$-tree $M_i$ of stage $\sigma$.

Let $M_i$ have $n_i$ vertices and 
$s_i$ $\ell$-subtrees.
Then
\begin{equation}
\label{PSubiEqn}
s_i\le n_i/2A_\ell(\sigma)
\le A_\ell(\sigma+1)/A_\ell(\sigma)
\end{equation}
where the first inequality 
uses the lower bound 
$2A_\ell(\sigma)$
on the size of 
an $\ell$-subtree
and the second inequality uses
the upper bound $2A_\ell(\sigma+1)$ on the size of 
an $\ell$-tree. 
The $(\ell-1)$-tree for $M_i$
has 
$s_i$ nodes. 
The
inductive assumption
shows the number of $add$s to form this  $(\ell-1)$-tree is
$\le 2s_i a_{\ell-1}(s_i)$.

Using the second inequality of \eqref{PSubiEqn} gives
$$
a_{\ell-1}(s_i) \le 
a_{\ell-1}(A_{\ell}(\sigma+1)/A_\ell(\sigma) )\le 
a_{\ell-1}(A_\ell(\sigma+1))=
a_{\ell-1}( A_{\ell-1}(A_\ell(\sigma)))=A_\ell(\sigma).$$
Using this and the first inequality of \eqref{PSubiEqn} shows the 
total number of $add$s  for all $(\ell-1)$-trees
of stage $\sigma$ is at most

\[\sum_i 
2s_i a_{\ell-1}(s_i) \le
2A_\ell(\sigma)\sum_i s_i \le 2A_\ell(\sigma)\sum_i n_i/2A_\ell(\sigma)  
=\sum_i n_i\le  n.
\]
This bound of $n$ $add$s per stage implies
$\le n a_\ell(n)$ recursive $add$s total.
This completes the induction.

Now consider the space.
There are initially $n$ nodes on level $L$. 
Additional nodes are only created in Case 1. 
The number of these nodes is obviously bounded by the number of 
\al. operations and so is $\le \eta$.
Theorem \ref{3.1Thm} 
shows the space used for incremental trees is 
proportional to $\eta$.
(As usual all space is allocated from one global array $S$ 
using Lemma \ref{SpaceDoublingLemma}.)
So \eqref{EtaEqn} implies the desired space bound.
\end{proof}

The remaining issue is how to choose the number of levels $\ell$.
Consider the usual case where bounds on 
$m$ and $n$ are known when the algorithm begins.
Take $\ell= \alpha(m,n)$. By definition
$a_{\alpha(m,n)}(n)\le 4\c{m/n}\le m/n+4$.
So the lemma implies the following.

\begin{theorem}
\label{3.2Thm}
A sequence of $\le m$ $nca$ and $link$ operations 
on a universe of $\le n$ nodes 
can be processed in time $O(m\alpha(m,n)+n)$ and space $O(m+n)$.
\hfill$\Box$\end{theorem}



Now we show that the same time bound can be achieved when
$m$ and $n$ are not known in advance.
In this setting
we allow the operation {\it make\_node}$(x)$ which creates a new node
$x$ in a singleton tree.
It is convenient to assume that such a node $x$ is not counted in $n$
until it is involved in a $link$ operation. (Since $\alpha(m,n)$ is increasing with $n$,  this can only make the desired time bound of Theorem \ref{3.2Thm}
stronger.) Since a $link$ increases $n$ by $\le 2$ we always have
\begin{equation*}
\label{mn2Eqn}
m\ge n/2.
\end{equation*}

We achieve the desired bound $ O(m\alpha(m,n)+n)$ using
a doubling strategy. The desired conclusion is not immediately clear
because of
two main difficulties: First,
$\alpha(m,n)$ is decreasing with $m$. So the time bound
for a
 $ca$ operation can decrease as the algorithm progresses. Second
the term $n a_\ell(n)$ in the bound of Lemma 
\ref{3.6Lemma} 
does not change in a predictable way.

\def\nlop.{$nca$/$link$}

We begin by describing our new procedure. It  uses
algorithm ${\cal A}_\ell$ where $\ell$ is
repeatedly modified. In precise terms
the sequence of operations is divided
into {\it periods}. The parameters $n$ and $m$ denote their values
at the start of a period. The period processes $nca$ and $link$ operations
using  algorithm
${\cal A}_\ell$ where $\ell=\alpha(m,n)$. (Note that the first period begins
with the execution of the first $link$, i.e., $m=1$, $n=2$, $\ell=1$.)
The period 
continues as long as the value of $\alpha$ remains in 
$\{\ell-1,\ell\}$. In other words we declare a new period whenever
$n'$ (the current number of $link$s), 
$m'$ (the current number of $link$s and
$nca$s), and
$\ell'=\alpha(m',n')$, have
$\ell'>\ell$ or $\ell'<\ell-1$.
The last period ends at the conclusion of the algorithm
($\alpha$ need not have changed).
In precise terms the algorithm is as follows.

\bigskip

{\narrower

{\parindent=0pt

Before executing the current $nca$/$link$ operation, 
update $n',m'$ and $\ell'$ to include that operation.
If $\ell'\in \{\ell-1,\ell\}$ then execute the operation.
Otherwise do the following:

\b

Set $n\gets n',\ m\gets m',\ \ell \gets \ell'$.
Reorganize the entire data structure to use algorithm ${\cal A}_{\ell}$.
Do this by making each current link tree $T$ an incremental tree
and placing it in the appropriate stage for $\A._\ell$. If $\ell>1$
add a corresponding node in stage 0 of level $\ell-1$. Finally execute
the current $nca$/$link$ operation.
}

}

\bigskip

This procedure clearly handles $link$s and $nca$s  correctly.
The time to start a new period is $O(n)$ for the new value of $n$.
(This includes the time to compute a new 
$ackermann$ table, find $\ell$, compute new incremental tree
tables and find the
stage for each incremental tree. All tables are computed
for the value $2n$, e.g., $ackermann$ stores all values
$A_i(j)\le 2n$.)

Now we prove the procedure 
achieves our goal. 
Note that the resource bounds 
of the following corollary 
are essentially the same as
Theorem \ref{3.2Thm}.

\begin{corollary}
\label{3.2Cor}
A sequence of $nca$ and $link$ operations
can be processed in time $O(m\alpha(m,n))$ and space $O(m)$.
Here $m$ is the number of $nca$s and $link$s,
$n$ is the number of $link$s, and neither is known in advance.
\end{corollary}

\def\bn.{\mathy{\overline n}}
\def\bm.{\mathy{\overline m}}
\def\bn.{\mathy{\bar n}}
\def\bm.{\mathy{\bar m}}

\begin{proof}
The bulk of the argument establishes the time bound.
We will charge processing time to the counts $n$ and $m$. To do this
define 
a {\em unit} of time to
be enough to pay for any constant amount of 
computing in the algorithm's time bound.

As in the algorithm, the analysis
uses $n$, $m$, and $\ell$ to
denote the values when the first \nlop. operation of
 the period is executed.
In addition we use $\bn.$ and $\bm.$ to denote the counts
up to and including  the last operation  of the period.
(So $\alpha(\bn.,\bm.)\in \{\ell-1,\ell\}$.)
We also use
 $n',m',\ell'$ to denote those values  for the first
operation of the next period. 
(So $m'=\bm.+1$. This holds for the last period by convention.
We also take $n'=\bn.$ for the last period.)
Note that  $m'-m$ is the number of operations in the period. 

The proof of the time bound consists of these three claims.

\claim 1 {The time for any period is at most $(m'-m)\ell + 12 \bm.$ units.}

\nclaim 2 {The total time from the start of the algorithm to the end of 
any period  but the last is at most $(12\ell'+50)\bm.$ units.
(As above, $\ell'=\alpha(m',n')$.)}

\claim 3 {When the last period ends, 
the total time for the entire algorithm is at most
$(12 \ell+62)\bm.$ units. 
The time bound of the corollary holds.}

\noindent
In Claim 1, all parameters (e.g., $\ell$)
are defined for the period under consideration.
In Claim 3, all parameters are defined for the last period.

\b

\noindent {\bf Proof of Claim 1.}
Lemma \ref{3.6Lemma}
shows the time for a period
is $O((m'-m)\ell +\bn. a_\ell(\bn.))$.
So we need only show $\bn. a_\ell(\bn.)\le 12\bm.$.

First observe 
\begin{equation}
\label{aEllEqn}
a_\ell(\bn.)\le 4 \c{\bm.\over \bn.}. 
\end{equation}
This is equivalent to $A_\ell( 4\c{\bm.\over \bn.})\ge \bn.$.
This inequality holds by definition if
$\alpha(\bm.,\bn.)=\ell$.
The other possibility is
 $\alpha(\bm.,\bn.)=\ell-1$.
But this also implies the same inequality, since
we have $A_\ell( 4\c{\bm.\over \bn.})\ge A_{\ell-1}( 4\c{\bm.\over \bn.})
\ge \bn.$.

The 
right-hand side of \eqref{aEllEqn}
is
$< 4 (\bm./\bn. +1) $.
Thus $\bn. a_\ell(\bn.) \le 4 (\bm. +\bn.)$.
Using $\bm.\ge \bn./2$ the last quantity is bounded by
$4\bm.+8\bm.=12\bm.$.
\ecproof

We prove Claims 2 and 3 by charging each $nca$/$link$ 
at most $\gamma$ time units, where
$\gamma$ is
$12 \ell' +50$ in Claim 2 and
$12 \ell+62$ in Claim 3.

\b

\noindent {\bf Claim 2 implies Claim 3.}
Claim 2 shows
the total time from the start of the algorithm to the beginning of the
last period is accounted for by
 a charge of $\gamma= 12\ell+50$, where $\ell=\alpha(m,n)$ is the value
used in the data structure of the last period.
(This holds {\em a fortiori} if the last period is actually the first period.)

Account for the time in the last period in two steps. First 
account for the term $(m'-m)\ell$ in Claim 1
 by charging
each $nca$/$link$ of the last period 
$\ell$ units. Since $\ell<\gamma$ every $nca$/$link$ 
is charged $\le \gamma$ units.
Next account for the term $12\bm.$ in Claim 1
by increasing $\gamma$ by 12, so the new charge is
$12 \ell+62$ units. This gives the first part of Claim 3.

The final value of $\alpha$ is $\ge
\ell-1$. So 
using the first part of Claim 3 and changing $\bm.$ to
the parameter $m$ of the corollary,
the time bound for the entire algorithm
is  $O(m (\alpha(m,n)+1) = O(m \alpha(m,n))$. Claim 3 is now completely proved.
\ecproof

\noindent {\bf Proof of Claim 2.}
Assume Claim 2 holds at the end of the previous period.
Now switch to the notation of the current period.
(The value $\ell'$ in Claim 2 becomes the parameter of the 
current period, $\ell$.) So each operation preceding the current
period is charged $\gamma= 12 \ell +50$ units.
Let $\ell'$ now
denote the value of $\alpha$ after the current period
ends. We wish to show the total time is accounted for
by charging every operation $\gamma'=12\ell' +50$ units.
Consider the two possibilities for $\ell'$.

\case {$\ell'\ge \ell+1$}
Account for the first term of Claim 1 by charging 
each $nca$/$link$ of the current
period $\ell$ units. Certainly $\ell< 12\ell+50$.
So now every $nca$/$link$ is 
charged $\le \gamma$ units.
Account for the second term by increasing 
$\gamma$ to $\gamma+12= 12(\ell+1)+50\le 12 \ell'+50$.
This gives Claim 2 for the current period.
(This case applies when the current period is the first period, since
$\ell=1$.)

\case {$\ell'\le \ell -2$}
If $m'\le 2m$ then \eqref{6Eqn} implies
$\ell'=\alpha(m',n')\ge \alpha(m,n)-1=\ell-1$.
So \[m'>2m.\]
\eqref{6Eqn} also implies 
$\ell'=\alpha(m',n')\ge \alpha(\bm.,\bn.)-1$
since 
$m'=\bm.+1\le 2\bm.$.
Using $\alpha(\bm.,\bn.)\ge \ell-1$
gives $\ell'=\alpha(m',n')\ge \ell-2$.
With the inequality assumed for  the current case, we get
\[\ell'=\ell-2.\]

Since $m'/2 >m$, we have $m'-m > m'/2$.
So we can account for the 
second term of 
Claim 1 by charging each $nca$/$link$ of this period
24 units ($24(m'/2)=12m'>12\bm.$).
The charge for the first term of Claim 1 is
$\ell$. So the total charge to each new time period is
$\ell +24= (\ell'+2)+24=\ell'+26$.

Each of the first $m-1$ operations is currently charged
$\gamma= 12\ell+50=12(\ell'+2)+50$.
Transfer 24 units to an operation of the current period.
(Permissible since $m'-m> m>m-1$.)
The first $m-1$ operations are now each charged
$\gamma'=12\ell'+50$. The operations of the current period are each charged
$\le (\ell'+26)+24<12\ell'+50=\gamma'$. 
So $\gamma'$ accounts for all the time so far.
\ecproof

Finally consider the space. Lemma \ref{3.6Lemma} shows the space for each period is $O(\bn. a_\ell(\bn.))$.
The argument of Claim 1 shows any period has
$\bn.a_\ell(\bn.) =O(\bm.)$.
Since $\bm.$ is at most the final value of $m$,
and the space for a period is always reused, the space bound follows.
\end{proof}

The multi-level method we have used can be applied to
achieve the same time and space bounds for several other problems.
As mentioned above, \cite{G85b} applies it to solve the
list splitting problem that arises in {\it expand} steps of Edmonds'
algorithm. The technique was  rediscovered by
Han La Poutr\'e: \cite{LaP} presents a
multi-level algorithm for the set merging problem (this application is
noted in \cite[p.\ 99]{G85b};
also a result similar to 
Corollary \ref{3.2Cor}  was independently arrived at [J.A. La Poutr\'e,
personal communication]. Other applications include
the static cocycle problem introduced in \cite{GS}, 
both for graphic matroids and the job scheduling matroid; the former
is useful for various problems involving spanning trees.


\or
\input intro
\input newed
\input bnotes

\input bmatch
\input code
\input strong
\fi

\ifcase 1 
\or

\section*{Acknowledgments}
The author thanks
an anonymous referee for a careful 
reading 
and many 
suggestions.

\clearpage
\setcounter{section}{0}
\renewcommand{\thesection}{\Alph{section}}
\renewcommand{\thetheorem}{\Alph{section}.\arabic{theorem}}

\setcounter{equation}{0}
\renewcommand{\theequation}{\Alph{section}.\arabic{equation}}
\section{Computing logarithms}
\label{LogAppendix}
We show how to compute $\f{\log_\beta r}$ for a given integer $r\in [1..cn^e]$
in time $O(1)$. 
Here $\beta=a/b$ is a fixed rational number for positive integers $a>b$,
$c$ and
$e$ are fixed integers, $c\ge 1,  e>1$.

Let $k=\f{\log_\beta n}$.
We precompute these values:

\bigskip

$\bullet$ $k, a^k, b^k$.

$\bullet$
a table $\ell[1.. n]$ with
$\ell[r] = \f{\log_\beta r}$ for  $r\in [1..n]$.

\bigskip

\noindent
We show the precomputation time is $O(n)$.

The following code precomputes the $\ell$ table:


\b

{\parindent=40pt

\def\For{{\bf for }}
\def\KwTo{{\bf to }}
\def\While{{\bf while }}
\def\do{{\bf do }}

$a'=1; \ b'=1; \ k=-1$

\For $r=1$ \KwTo $n$ \do

{\hi
 
\While{$r\ge a'/b'$} \do


{\hi

$a'=aa';\ b'=bb';\  k=k+1$

}

$\ell[r]=k$

}
}

\b

\noindent 
On exit $k$ is the desired value $\f{\log_\beta n}$ and 
the desired values $a^k$ and $b^k$ are given by
$a'/a$ and $b'/b$ respectively. It is clear that the time is $O(n)$.

Now we give the algorithm to compute
$\f{\log_\beta r}$ for a given integer $r\in [1..cn^e]$.
Let $h$ be the unique integer satisfying
\[\beta^{hk} \le r <\beta^{(h+1)k}.\]
So
\begin{equation}
\label{LogEqn}
\log_{\beta}r=hk + \log_\beta (r/{\beta^{hk}}).
\end{equation}
Taking floors gives the desired value.

We find $h$ by testing successive values $\beta^{hk}$.
The desired $h$ is at most $e + (e+\log_\beta c) /k=O(1)$.
This follows for $r\le c n^e$ since
$c=\beta^{\log_\beta c}$ and
$n<\beta^{k+1}$ implies $n^e<\beta^{ke+e}$.
Using the values $a^k,b^k$
the time is $O(e)=O(1)$. 

The desired floor of the logarithmic term in \eqref{LogEqn} is
$\f{\log_\beta \f{r/{\beta^{hk}}}}$. This corresponds to an entry in the
$\ell$ table since $r/{\beta^{hk}}<\beta^{k}\le n$. The desired entry is
found as $\ell[rb^{hk}/a^{hk}]$ (since division is truncating).
Again the time is $O(1)$.

\section{Simple inequalities for Ackermann's function}
\label{AckAppendix}

\noindent
{\em Proof of \eqref{4Eqn}, $A_i(j+1) \ge 2A_i(j)$ for $i,j \ge 1$}:

First note the trivial inequality $2^i\ge 2i$ for $i\ge 1$.
Also for every $i\ge 1$, $A_i(2)=4$.

Next we show $A_i(j)\ge 2^j$ by induction on $i$, 
with the inductive step inducting on $j$. 
The base case $j=2$ of the inductive step
is $A_i(2)=4=2^2$.
For the inductive step
\[A_i(j) = A_{i-1}(A_i(j-1) )\ge 2^{A_i(j-1) }\ge 2^{2^{j-1}} \ge
2\cdot 2^{j-1}=2^j.\]
Now
\eqref{4Eqn} itself follows from the 
first $\ge$ relation displayed above
and $2^{A_i(j-1) }\ge 2A_i(j-1)$.

\b

\noindent
{\em Proof of \eqref{5Eqn}, $A_{i+1}(j) \ge A_i(2j)$ for $i \ge1,j \ge 4$}:

Note that \eqref{5Eqn} needn't hold for $j<4$:
Recalling that $A_2$ is superexponentiation $A_2(j)=2\uparrow j$,
$A_2(3)=16<64=A_1(6)$.
However \eqref{5Eqn} holds for $i=1,j=4$:
$A_2(4)=2^{16}>2^8=A_1(8)$. 
In general for $i\ge 1$, 
\[A_{i+1}(4)= A_i(A_{i+1}(3))=A_i(A_i(A_{i+1}(2)))=
A_i(A_i(4)).\]
Also $A_i$ is an increasing function by
\eqref{4Eqn}.

We prove \eqref{5Eqn} by induction on $i$. 
For the base case,
$A_{i+1}(4)= A_i(A_i(4))\ge A_i(2^4)>A_i(8)$.
For the inductive step,
$A_{i+1}(j+1)
= A_{i}(A_{i+1}(j) )$. The argument to $A_i$ is
$A_{i+1}(j) 
> A_{i}(2j) \ge 2^{2j}
\ge 2(2j)\ge 2j+2$.

\b

\noindent
{\em
Proof of \eqref{6Eqn}, $\alpha(m',n')\ge\alpha(m,n)-1$ for $m'\le 2m, n'\ge n$}:

Let $i=\alpha(m,n)$. 
We wish to show $\alpha(m',n')\ge i-1$, i.e.,
$A_{i-2}(4 \c{ {m' \over n'} } ) <n'$. This follows since
\[
A_{i-2}(4 \c{ {m' \over n'} } ) \le
A_{i-2}(8\c{ m\over n })\le
   A_{i-1}(4\c{ m \over n })<n\le n'.\] 
The first inequality uses 
$\c{m'/n'}\le \c{2m/n}\le 2\c{m/n}$. The second uses
\eqref{5Eqn}, which applies since 
$4\c{ m \over n }\ge 4$. 
(A
slightly more involved calculation shows
$\alpha(m',n')\le\alpha(m,n)+2$
when  $m'\ge m, n'\le 2n $ but we do not use this fact.)


\fi 

\begin{thebibliography}{99}

\def\referencesheading{0}
\ifcase\referencesheading
\or
\bigskip\penalty-2000%
\noindent{\bf References \hfill\bigskip}
\parindent=0pt
\or
\line{\quad}\penalty-2000%
\noindent{\twelvebf References \hfill}
\medskip
\parindent=0pt
\fi
%
\font\ninerm=cmr9
\font\ninebf=cmbx9
\font\nineit=cmti9        
\font\ninei=cmmi9
\font\ninesy=cmsy9
\def\ninepoint{%
   \def\rm{\ninerm}\def\bf{\ninebf}%
   \def\it{\nineit}\def\smc{\ninerm}\baselineskip=11pt\rm%
        \textfont0=\ninerm \scriptfont0=\sevenrm
	\textfont1=\ninei \scriptfont1=\seveni
	\textfont2=\ninesy \scriptfont2=\sevensy
	\textfont3=\tenex \scriptfont3=\tenex
}
\def\rsize{39pt}
\def\r #1]{\medskip%
\hangafter=-10\hangindent=\rsize%
\hskip0pt  \llap{\hbox to \rsize{#1]\hfill}}\ignorespaces}
%
%
\def\al #1,{{\it Algorithmica, #1,}}
\def\comb #1,{{\it Combinatorica, #1,}}
\def\cu,{Comp.\ Sci.\ Dept., Univ.\ Colorado, Boulder, CO,}
\def\focs #1,{{\it Proc.\ #1 Annual Symp.\ on Found.\ of Comp.\ Sci.,}}
\def\ipl #1,{{\it Inf.\ Proc.\ Letters, #1,}}
\def\ja #1,{{\it J.\ Algorithms,  #1,}}
\def\jacm #1,{{\it J.\ ACM,  #1,}}
\def\jcss #1,{{\it J.\ Comp.\ and System Sci., #1,}}
\def\mprog #1,{{\it Math.\ Programming,  #1,}}
\def\mprogb #1,{{\it Math.\ Programming B,  #1,}}
\def\net #1,{{\it Networks, #1,}}
\def\phd{Ph.\ D.\ Dissertation}
\def\sicomp #1,{{\it SIAM J.\ Comput.,  #1,}}
\def\siad #1,{{\it SIAM J.\ Alg.\ Disc.\ Meth., #1,}}
\def\sidm #1,{{\it SIAM J.\ Disc.\ Math., #1,}}
\def\soda #1,{{\it Proc.\ #1 Annual ACM-SIAM Symp.\ on Disc.\ Algorithms,}} 
\def\stoc #1,{{\it Proc.\ #1 Annual ACM Symp.\ on Theory of Comp.,}}
\def\tr {Tech. Rept.\ }
%
\def\pp#1-#2.{pp.\ #1--#2.}
\def\spp#1-#2;{pp.\ #1--#2;}

%
\def\nrlq #1,{{\it Naval Res.\ Logist.\ Quart., #1,}}
%

\bibitem{AHU}
A.V. Aho, J.E. Hopcroft, and J.D. Ullman,
"On finding lowest common ancestors in trees'',
\sicomp 5, 1976, \pp 115-132.


\bibitem{BV}
O. Berkman and U. Vishkin, 
"Recursive star-tree parallel data 
structure", \sicomp 22, 2, 1993, \pp 221-242.


\bibitem{CH}
R. Cole and R. Hariharan, 
``Dynamic LCA queries on trees'',
\sicomp 34, 4, 2005, \pp 894-923.


\bibitem{CLRS}
T.H.~Cormen, C.E.~Leiserson, R.L.~Rivest and C.~Stein,
{\em Introduction to Algorithms}, 2nd Ed.,
McGraw-Hill, NY, 2001.

\bibitem{Ed}
J. Edmonds, ``Maximum matching and a polyhedron with 0,1-vertices'', 
{\it  J.\ Res.\ Nat.\ Bur.\ Standards 69B}, 1965, \pp 125-130.


\bibitem{FS}
M.L. Fredman and M.E. Saks, ``The cell probe complexity of dynamic data
structures'', 
\stoc 21st, 1989, \pp 345-354.

\bibitem{G85b} 
H.N. Gabow,
``A scaling algorithm for weighted matching on general graphs'',
\focs 26th, 1985, \pp 90-100.

\bibitem{G90}
H.N. Gabow, 
``Data structures for weighted matching and nearest common
ancestors with linking'',
\soda 1st, 1990, \pp 434-443. 

\bibitem{G17}
H.N. Gabow, 
"Data structures for weighted matching 
and extensions to $b$-matching and $f$-factors'',
2016, currently submitted for publication.

\bibitem{GS}
H.N. Gabow and M. Stallmann, 
``Efficient algorithms for graphic matroid intersection and parity'',
{\it Automata, Languages and Programming: 12th Colloquium,}  
{\it Lecture Notes in Computer Science 194},
W. Brauer, ed., Springer-Verlag, 1985, \pp 210-220.

\bibitem{Gus}
D.~Gusfield, {\em Algorithms on Strings, Trees, and Sequences},
Cambridge University Press, NY, 1999, Ch. 9.



\bibitem{HT}
D. Harel and R.E. Tarjan, ``Fast algorithms for finding nearest common 
ancestors'', \sicomp 13, 2, 1984, \pp 338-355. 

\bibitem{LaP}
J.A. La Poutr\'e,
``New techniques for the union-find problem'',
\soda First, 1990, \pp 54-63.

\bibitem{S}
A.~Schrijver,
{\it Combinatorial Optimization: Polyhedra and Efficiency},
Springer, NY, 2003, Ch.~26.

\bibitem{SlT} 
D.D. Sleator and R.E. Tarjan,  ``A data structure for dynamic trees'', 
\jcss 26, 1983, \pp 362-391.

\bibitem{Smy}
B.~Smyth, {\em Computing Patterns in Strings},
Pearson Education Limited, Harlow, England, 2003.

\bibitem{SV}
B. Schieber and U. Vishkin, ``On finding lowest common 
ancestors: simplification and parallelization'',
\sicomp 17, 6, 1988, \pp 1253-1262. 

\bibitem{T79}
R.E. Tarjan, ``Applications of path compression on balanced trees'', 
\jacm 26, 4, 1979, \pp 690-715.


\bibitem{T83} R.E. Tarjan,
{\it Data Structures and Network Algorithms,}
SIAM, Philadelphia, PA., 1983.


\end{thebibliography}
\end{document}